\documentclass[a4paper,11pt]{article}

\usepackage{jcappub} 

\usepackage[T1]{fontenc} 

\usepackage{pdflscape}  

\usepackage{multirow}
\usepackage{booktabs}      
\usepackage{array}         %
\usepackage{amsmath}       

\usepackage{comment}
\usepackage{url}
\usepackage{hyperref}  
\usepackage{etoolbox} 
\usepackage{latexsym}
\usepackage{graphics}

\newcommand{\bi}{\begin{itemize}}
	\newcommand{\ei}{\end{itemize}}

\newcommand{\bnum}{\begin{enumerate}}
	\newcommand{\enum}{\end{enumerate}}

\newcommand{\bvb}{\begin{verbatim}}

\newcommand{\ba}{\begin{eqnarray}}
	\newcommand{\ea}{\end{eqnarray}}

\newcommand{\eq}[1]{(\ref{#1})}

\newcommand{\be }{\begin{equation}}
	\newcommand{\ee}{\end{equation}}

\title{\boldmath Model-independent test of the cosmic distance duality relation with recent observational data}

\author{Xing Wu}

\affiliation{Department of Physics, North University of China \\3 Xueyuan Road, Taiyuan, Shanxi 030051, China}

\emailAdd{xwu@nuc.edu.cn}

\abstract{We test the cosmic distance duality relation (CDDR) using two model-independent methods. Method I is based on the PAge parametrization, which characterizes the expansion history in terms of the cosmic age. Parametrizations of possible CDDR violations are constrained using observational data from Type Ia supernovae (SN), baryon acoustic oscillations (BAO), cosmic chronometers, and gamma-ray bursts (GRB), including the latest PantheonPlus and DES Dovekie SN samples and DESI DR2 BAO data. The results support the validity of the CDDR within $1\sigma$. Different combinations of data sets are further explored to assess the impact of various probes and calibration choices, demonstrating the robustness of this conclusion. Although GRB data extend to higher redshifts, their constraining power is significantly weaker than that of the other low-redshift probes. The PantheonPlus and DES Dovekie samples yield consistent results. Method II uses a non-parametric Gaussian process reconstruction of the luminosity distance from SN data, combined with BAO measurements to construct the observed CDDR violation and constrain its parametrizations. The results are consistent with those from Method I, and we find no evidence for a violation of the CDDR.}

\begin{document}
	\maketitle
	\flushbottom

\section{Introduction}
 	
The concept of distance is a fundamental pillar of cosmology. Various definitions of distance \cite{Weinberg:1972kfs,astro-ph/9905116} are used in cosmology, depending on the observational method. The cosmic distance duality relation (CDDR, also known as the Etherington distance duality \cite{2007GReGr..39.1047E}), $d_L=(1+z)^2d_A$, relates two widely used distances: the luminosity distance $d_L$ based on the observed luminosity of an object of known intrinsic brightness, and the angular diameter distance $d_A$ based on the angular size of an object of known length scale. This relation is valid under the assumptions that gravity is described by a metric theory, the observed photons follow null trajectories, and their number is conserved. Any violation of the CDDR would signal a departure from the above assumptions, e.g., new theories of gravity beyond general relativity  \cite{1703.10871,2104.01209}, coupling between photons and exotic particles beyond the standard particle physics  \cite{hep-ph/0111311,astro-ph/0311495, 0711.2966}, or simply cosmic opacity due to some unaccounted sources  \cite{1004.2053,1512.01861}. Testing the validity of the CDDR is therefore of fundamental importance in cosmology.

The violation of the CDDR is commonly written as
\be \label{eta z def}
\eta(z)=\frac{d_L}{(1+z)^2d_A}.
\ee
Since the early works \cite{astro-ph/0312443, astro-ph/0405620}, 
numerous studies have tested the CDDR using various methodologies and different observational data with continuously improved quality. Most approaches construct $\eta_\text{obs}$, which can then be used for parametric or non-parametric reconstruction of $\eta(z)$ \cite{2110.00927, 2507.13811, 2512.06454}. This requires both $d_L$ and $d_A$ be obtained at the same redshifts, or at least within the same range. For instance, $d_L$ is typically obtained from Type Ia supernova (SN) data and $d_A$ from baryon acoustic oscillations (BAO) data. To match the redshifts of SN with those of BAO, one may select SN data very close to the BAO redshifts according to a certain criterion (e.g., the redshift difference $|z_\text{SN}-z_\text{BAO}|<0.005$), as done in early works \cite{1005.4458,1102.1065,1104.4942}. Alternatively, a binning method was  proposed in  \cite{1104.2833} to align $d_L$ and $d_A$ at the same redshifts to test the CDDR \cite{Wu:2015ixa, 2407.12250}. More recently, one can benefit from the advancement of machine learning techniques and obtain $d_L(z)$ from the SN data using data-driven methods, such as Gaussian process (GP),\footnote{
	In general, GP is used to solve two types of problems: regression and classification. In this paper, GP specifically denotes Gaussian process regression.
} artificial neural network (ANN) \cite{2507.13811,2512.06454}, or the so-called neural kernel Gaussian process regression \cite{2508.07040}, and then extract the $d_L$ values at the BAO redshifts to produce the corresponding $\eta_\text{obs}$. With $\eta_\text{obs}$ available, one can either constrain a parametrized form of $\eta(z)$ (e.g., \cite{1003.5906,1102.1065}), or perform a non-parametric construction of $\eta(z)$ (e.g., \cite{2506.17926,2509.07848}). Besides BAO data, $\eta_\text{obs}$ were also constructed with $d_A$ from galaxy cluster data \cite{1005.4458} and megamaser data \cite{2507.11518}.

Another method is to assume a specific parametrization of $\eta(z)$ and constrain the CDDR violation parameter jointly with the cosmological model parameters. For example, in $\Lambda$CDM and $w_0w_a$CDM, $\eta(z)$ was constrained together with the cosmological parameters in  \cite{2504.10464}.    
Alternatively, based on some cosmological model-independent parametrizations of $H(z)$, such as the Pad\'e approximation expressed in terms of cosmographic parameters \cite{2408.13390} and B\'ezier polynomials \cite{2509.09247}, $\eta(z)$ can be constrained together with the parameters of $H(z)$.

Most studies of the CDDR use low redshift observational data such as SN and BAO, which are currently limited to $z\lesssim 2.3$. As a result, possible evolution of the CDDR violation is less constrained by higher-redshift data. Given that the cosmic microwave background (CMB)  measurements have yielded precise results about the Universe at very high redshifts $z\sim 10^3$, more probes exploring the intermediate redshift range can provide important information about the cosmic expansion history. To use data extending to higher redshifts, one can use strong gravitational lensing (SGL) data (e.g.,  \cite{1509.07649,1809.09845,1906.04107}), which can reach $z\sim3$ to $4$, to test the CDDR \cite{1511.01318,1710.10929}. In the future, strongly lensed gravitational wave data \cite{2603.23373} may extend to even higher redshifts and provide more constraints. In addition, quasars \cite{1902.01988} and gamma-ray bursts (GRB) \cite{2509.09247,2512.06454} are also used to extend the CDDR test to higher redshift.

Besides being tested, the CDDR itself is also useful in many other contexts, e.g.,  testing the galaxy cluster gas mass density profile \cite{1601.00409}, determining the Hubble constant using the inverse distance ladder  \cite{1811.02376, 2406.05049}, and studying the Hubble tension  \cite{2504.10464}.

In this paper, we use two methods, denoted as Method I and Method II, to test the CDDR in a model-independent manner, in the sense that the analysis is not restricted to any particular late-time cosmological model. Common model-independent approaches are based on cosmography  \cite{gr-qc/0309109}, Pad\'e approximation  \cite{1309.3215,1312.1117}, B\'ezier curve  \cite{1811.08934}, or data-driven methods such as GP \cite{1204.2832} and ANN \cite{1910.03636}. In Method I, we use the PAge parametrization \cite{2001.06926}, a model-independent description characterized by the cosmic age, to constrain several parametric forms of $\eta(z)$ jointly with cosmological parameters, using observational data of SN, BAO and Cosmic Chronometers (CC), combined with GRB data to extend the redshift range to $z\sim 8$. Unlike other parametrizations such as the cosmographic expansion or the B\'ezier curve method, the PAge parametrization with only three parameters provides a good approximation to a large class of cosmological models up to high redshifts. For example, as shown in  \cite{2107.13286}, the deviations of $H(z)$ and $d_L(z)$ in the PAge parametrization from those in the $\Lambda$CDM model remain small over a broad redshift range up to $z\sim 10^4$. In Method II, we use GP to reconstruct $d_L$ directly from SN data, and then construct $\eta_\text{obs}$ values, which are used to constrain parametrizations of $\eta(z)$. Note that the GRB data cannot be used in Method II, since there are currently no observational data that provide $d_A$ up to the same high-redshift range as GRB.  

The paper is organized as follows. The PAge method is introduced in Section \ref{sec:PAge}. The observational data used in this work are presented in Section \ref{sec:data}. Analyses using Method I and Method II are performed in Sections \ref{sec:Method I} and \ref{sec:Method II}, respectively. Conclusions are presented in the last section.

\section{PAge parametrization}
\label{sec:PAge}
 
Model-independent methods studying the late-time expansion history of the universe typically reply a certain parametrization of the Hubble parameter or distances such as the luminosity distance and angular diameter distance.

The PAge parametrization, characterized by the cosmic age, was proposed in  \cite{2001.06926} (see also \cite{2008.00487}), originally introduced to confirm the current cosmic acceleration based on the observation that the cosmic age $t_0\geq 12$Gyr, without resorting to supernovae. The parametrization is motivated as follows. In a matter-dominated universe, one has $E(z)=\sqrt{\Omega_m(1+z)^3}$, corresponding to the so-called Einstein-de Sitter universe. It can be shown that the product of the Hubble parameter $H(t)$ and the cosmic age $t$ is fixed to $Ht=2/3$. Since our universe is indeed dominated by matter over an extended period, and $Ht$ evolves slowly and smoothly, the actual expansion history of the universe can be approximated as evolving from the Einstein-de Sitter universe at early times $t\approx 0$, and gradually deviating from it at later times up to the present epoch $t=t_0$. Accordingly, $Ht$ can be approximated by a Taylor expansion around small $t$ to uadratic order 
\be 
Ht\approx\frac{2}{3}+A\frac{t}{t_0}+B\left(\frac{t}{t_0}\right)^2,
\ee 
where $A$ and $B$ can be determined from $t_0$ and $q_0$, the current value of the deceleration parameter $q(t)\equiv-\ddot a a/\dot a^2$. The Hubble parameter $H(t)$ in the PAge parametrization can be determined from
\be\label{Huang page}
E\equiv H(t)/H_0=1+\frac{2}{3}\left(1-\eta\frac{H_0t}{p_\text{age}}\right)\left(\frac{1}{H_0 t}-\frac{1}{p_\text{age}}\right)~,
\ee
where  
\be\label{p eta in page}
p_\text{age}\equiv H_0t_0,\ \ \ \eta\equiv 1-\frac{3}{2}p_\text{age}^2(1+q_0)~.
\ee
Here $p_\text{age}$ corresponds to the current age of the universe in units of $H_0^{-1}$ and $\eta$ characterizes the deviation from the Einstein-de Sitter universe with $\eta=0$.\footnote{
	This parameter is unrelated to the CDDR violation parametrization $\eta(z)$ or its parameter $\eta_0$. The notation follows conventional usage in the literature and should be distinguished by context.
}

PAge provides a sufficiently accurate global approximation to a large class of late-time cosmological models over a broad redshift range, which is probed by most current observations such as SN, BAO, CC, and GRB. The errors of the approximation to typical cosmological models  are well controlled \cite{2008.00487, 2012.02474, 2107.13286, 2108.03959}. For example, the relative errors in $d_A$ and $d_L$ in the PAge approximation, compared to models such as $\Lambda$CDM, $w$CDM, and the CPL parametrization, are below $0.01$ for $z<10$ \cite{2012.02474, 2108.03959, 2202.12214}. This is sufficient for our analysis and we therefore do not consider the improved cosmic age expansion proposed in \cite{2108.03959}, which is slightly more accurate at the expense of introducing an additional parameter. The correspondence between the PAge parameters $(p_\text{aga},\eta)$ and those of other models can be found in  \cite{2008.00487, 2108.03959}. Observational constraints on the PAge parametrization can also be found in \cite{2505.22369,2510.26355}

By introducing the dimensionless cosmic age $\tau=H_0t$, Equation~\eq{Huang page} can be rewritten as 
\be \label{PAge in tau}
\frac{H(\tau)}{H_0}=1+\frac{2}{3}\left(1-\eta\frac{\tau}{p_\text{age}}\right)\left(\frac{1}{\tau}-\frac{1}{p_\text{age}}\right),
\ee
with $\tau_0=p_\text{age}$ denoting the present dimensionless cosmic age. 
Furthermore, $\tau$ can be regarded as a function of redshift $z$. Indeed, using
\be \label{H dz dt}
H=-\frac{dz}{dt}\frac{1}{1+z},
\ee
Equation~\eq{PAge in tau} can be recast as a differential equation for $\tau$, which leads to the solution
\be\label{PAge z and t}
1+z=\left(\frac{p_\text{age}}{\tau}\right)^{2/3} e^{\frac{1}{3}(1-\frac{\tau}{p_\text{age}})(3p+\eta\frac{\tau}{p_\text{age}}-\eta-2)}.
\ee
For a given $z$, the corresponding $\tau=\tau_z$ is determined by Equation~\eq{PAge z and t}. Inserting $\tau_z$ into Equation~\eq{PAge in tau} yields $H/H_0$ as a function of $z$, which is convenient for numerical calculation. In particular, the angular diameter distance is (with spatial flatness assumed in this work)
\be 
d_A(z)=\frac{c}{H_0(1+z)}\int^z_0\frac{dz}{E(z)}.
\ee
From Equation~\eq{H dz dt}, it follows that
\be 
\int^z_0\frac{dz}{E(z)}=\int_{\tau_z}^{\tau_0}(1+z)d\tau,
\ee
which allows $d_A$ to be written as a function of $\tau_z$
\be 
d_A(\tau_z)=\frac{c}{H_0(1+z)}\int_{\tau_z}^{\tau_0}(1+z)d\tau.
\ee
where $(1+z)$ is understood as a function of $\tau$, given by Equation~\eq{PAge z and t}.

\section{Observational data}
\label{sec:data}

The data used in this work include SN, BAO, cosmic chronometer (CC), and GRB. The SN and GRB data are related to $d_L$, and the BAO data are related to $d_A$. The CC data are introduced to provide calibration, as will be discussed in the next section.

\subsection{Type Ia supernovae}

\label{subsection:SN}

Type Ia supernovae (SNe), as one of the well-established cosmological probes and standard candles, typically provide the $B$-band apparent magnitude $m_B$, which can be standardized to $m_B^{corr}$ by considering corrections from light curve stretch, color, selection biases, and the correlation with host-galaxy properties such as mass and color. Then the distance modulus $\mu$ and the luminosity distance $d_L$ are given by
\be \label{SN mu }
\mu=m_B^{corr}-M_B=5\log_{10} \frac{d_L}{\text{Mpc}}+25, 
\ee
where $M_B$ is the absolute magnitude. When considering a possible CDDR violation parametrized by $\eta(z)$, the luminosity distance becomes
\be \label{dL with eta}
d_L(z)=\eta(z)(1+z)\frac{c}{H_0}\int^z_0\frac{dz}{E(z)}.
\ee

We use the PantheonPlus SN data set \cite{2202.04077}, which contains data from 1701 light curves of 1550 spectroscopically identified distinct type Ia SNe within the range $0.001< z<2.26$. Among these, 77 data points (from 42 SNe denoted as ``calibrators'' in 37 Cepheid host galaxies) are used in SH0ES  \cite{2112.04510} to calibrate the SN distances by Cepheid variables, yielding an absolute magnitude $M_B=-19.253\pm 0.027$. The covariance matrix of the PantheonPlus SN data  \footnote{\url{https://github.com/PantheonPlusSH0ES/DataRelease}} contains not only the covariance between SNe, but also the Cepheid host covariance  \cite{2202.04077}.\footnote{This is tailored to the likelihood function used in Eqs.~(14) and (15) of \cite{2202.04077}. But this likelihood function is not used in this work.}

For the purpose of cosmological parameter estimation, following  \cite{2202.04077} (see also \cite{2505.22369,2407.17252}), we exclude 111 data points of nearby SNe with $z <0.01$ to reduce the impact of peculiar velocities at low redshifts. This leaves 1590 data points, from which we further remove 10 of the 77 SH0ES calibrators with $z<0.01$, resulting in 1580 SN data points used in this work (see also \cite{2212.07917}).

We also use the most recent Dark Energy Survey (DES) Dovekie data set\footnote{
	\url{https://github.com/des-science/DES-SN5YR}}
\cite{2511.07517}, which is based on a complete reanalysis of the DES5yr sample \cite{2401.02929} with an improved calibration solution  \cite{2506.05471} and a fixed host galaxy dust color law to better control systematics. The data set contains 1820 SNe with $0.025<z<1.15$, including 1623 photometrically confirmed type Ia SNe from DES and 197 spectroscopically confirmed low-redshift type Ia SNe from external surveys. For the photometrically confirmed SNe, in order to take account of the possible non-type Ia SN contamination, the uncertainties of the distances from DES are renormalized by the Bayesian Estimation Applied to Multiple Species (BEAMS) probability of being type Ia \cite{2306.05819, 2401.02945,2511.07517}. Consequently, the uncertainties of likely contaminants are greatly enlarged, thereby significantly reducing their influence in cosmological analyses.

PantheonPlus and DES Dovekie are the two most recent data sets, providing high-quality SN Ia data with different number densities over the redshift range. They are obtained by different observational methods and processed using different techniques, and thus provide independent measurements of $d_L$.

We use the following $\chi^2$ for the SN data
\be 
\chi_\text{SN}^2=\sum_{ij}\Delta\mu_i (C^{-1})_{ij}\Delta\mu_j,
\ee
where $C$ is the covariance matrix including full statistical and systematic uncertainties,  and $\Delta\mu_i=\mu^\text{obs}_i-\mu_i^\text{th}$ is the difference between observed and theoretical distance moduli. For $\mu^\text{obs}$, we use the $m_B^{corr}$ data in both PantheonPlus and DES Dovekie, \footnote{
	Unlike PantheonPlus, the $m_B^{corr}$ values are not directly provided by DES Dovekie, but they can be easily recovered via the available data from its github repository.
} 
with $M_B$ treated as a parameter to be calibrated. This $\chi^2_\text{SN}$ depends on the CDDR violation parameter $\eta_0$, in addition to the PAge parameters $p_\text{age}$, $\eta$, and $H_0$.

\subsection{Baryon Acoustic Oscillations}

BAO are imprints on the late-time large scale structure left by the  propagation of fluctuations in the plasma before recombination in the early universe. They are detected as a bump in the two-point correlation function (2PCF) of matter tracers (e.g., galaxies or quasars), or equivalently, as a set of characteristic oscillations in the power spectrum, which is the Fourier transform of the 2PCF. The characteristic scale of the BAO, which serves as a standard ruler, is given by the comoving sound horizon 
\be 
r_d=\int_{z_d}^\infty\frac{c_s(z) dz}{H(z)},
\ee
where $c_s$ is the speed of sound, and $z_d\approx1060$ \cite{1807.06209} is the redshift at which the acoustic waves freeze out at the end of the drag epoch. BAO measurements probe the standard ruler along the transverse and radial directions. The transverse BAO provides 
\be 
\frac{d_M(z)}{r_d}=\frac{(1+z)d_A}{r_d},
\ee
where $d_M$ is the transverse comoving distance. The line-of-sight BAO yields the Hubble distance $d_H$ as
\be\label{eq:relative H measurements}
\frac{d_H(z)}{r_d}= \frac{c}{H(z)r_d}.
\ee
In redshift ranges with low signal-to-noise ratio, BAO measurements produce the spherically-averaged distance  
\be
\frac{d_V(z)}{r_d} =\frac{[zd_M^2(z)d_H(z)]^{1/3}}{r_d} . 
\ee
In practice, one can use the CMB or big bang nucleosynthesis (BBN) results of $r_d$ to obtain absolute measurements of $d_M$ and $d_H$, as commonly done in the inverse distance ladder analysis \cite{1811.02376,2406.05049}. In the following, when performing the inverse distance ladder style analysis, we adopt the Planck 2018 \cite{1807.06209} CMB (TT,TE,EE+lowE) result $r_d=147.05\pm 0.30$ Mpc. We also  use the relative BAO measurements and leave $r_d$ as a free parameter.

For the BAO data, we use DESI DR2 \cite{2503.14738}, which include the relative measurements and the covariance matrix.\footnote{
	\url{https://github.com/CobayaSampler/bao_data/tree/master/desi_bao_dr2}} The $\chi^2$ for BAO is constructed as
\be 
\chi^2_\text{BAO}=\sum_{ij} \Delta D_iC^{-1}_{ij}\Delta D_j,
\ee
where $\Delta D_i$ denotes the difference between theory and data of the relative distances mentioned above, and $C$ is the covariance matrix.

\subsection{Cosmic Chronometers}

The observed Hubble data (OHD) $H(z)$ can be obtained using the CC method, which is based on the differential age of passive galaxies \cite{astro-ph/0106145}, and is independent of cosmological models. Note that the line-of-sight BAO measurements can also provide relative $H(z)$ measurements as in Equation (\ref{eq:relative H measurements}), but such measurements are regarded as BAO data in this work.

A commonly used compilation in the literature includes 32 CC data points, as compiled in \cite{2201.07241, 2205.13247}. In this work, we use 33 CC data points (see also \cite{2502.11443}), including one additional measurement at $z=1.26$ from  \cite{2305.16387}. Moreover, the data point at $z=0.75$ from  \cite{2110.04304} is replaced with that at $z=0.8$ from \cite{2205.05701}. This is because the two OHD measurements are obtained from different subsamples of a common parent sample, i.e., the LEGA-C DR2 survey \cite{1809.08236}, using different methods. Compared to the idealized model used in  \cite{2110.04304}, the method used in  \cite{2205.05701} is improved by using a more realistic star formation history model, which also allows for a larger subsample of CC and thus a smaller statistical error. Since the two measurements are covariant, they should not be used simultaneously and we therefore adopt only the result of \cite{2205.05701}. 
In addition, as noted in the appendix of  \cite{2502.11443}, the commonly used value $H(0.09)=69\pm12$ km/s/Mpc from  \cite{astro-ph/0302560} actually corresponds to  the inferred $H_0$ value in that paper, and should be corrected to $H(0.09)=70.7\pm12.3$ km/s/Mpc.

A method was proposed in \cite{2003.07362} to estimate the covariance between 15 of the 33 OHD points, yielding a full $15\times 15$ covariance matrix.\footnote{\url{https://gitlab.com/mmoresco/CCcovariance}}  However, this method cannot be directly applied to the remaining 18 OHD points, and their correlations are not available. Some studies simply used the $15\times 15$ covariance matrix together with the uncorrelated uncertainties for the remaining data \cite{2505.22369}. We adopt the same approach. Note that some studies have attempted to expand the existing $15\times 15$ covariance matrix to a larger size using numerical simulations and data-driven methods, as discussed in \cite{2208.13700, 2502.11443}. 

For the CC data, we use
\be 
\chi^2_\text{CC}=\sum_{ij} \Delta H_iC^{-1}_{ij}\Delta H_j.
\ee
where $\Delta H_i$ denotes the difference between the theoretical prediction and the observed value of $H(z)$, and $C$ denotes the covariance matrix.

\subsection{Gamma-Ray Bursts}

The GRB\footnote{Throughout the paper, GRB refer specifically to long GRB, as short GRB are not standardizable and cannot be directly used to constrain cosmological parameters.} 
data are used to extend the redshift range to much higher values. Although GRB are not standard candles, they can be standardized using a set of empirical laws, such as the Amati relation \cite{astro-ph/0205230} and the Combo relation \cite{1508.05898} (see \cite{2201.07241} for a recent review). Here, we focus on the Amati relation
\be\label{Amati relation}
y=a +bx,
\ee
where $y=\log_{10}(E_\text{iso}/\text{erg})$, $x=\log_{10}(E_\text{p,i}/300\text{keV})$,
$E_\text{iso}$ is the isotropic energy, $E_\text{p,i}$ is the rest-frame spectral peak energy, and $a$ and $b$ are parameters of this linear relation to be fitted. The luminosity distance $d_L$ enters through $E_\text{iso}$ via
\be\label{E iso dL S bolo}
E_\text{iso}=4\pi d_L^2 \frac{S_\text{bolo}}{1+z},
\ee
where $S_\text{bolo}$ is the observed bolometric fluence. Calibrating the Amati relation using $d_L$ from a given cosmological model would lead to the so-called circularity problem when using this relation to constrain the cosmological model. To circumvent this issue, one can either fit the Amati relation and the cosmological model together \cite{0805.0377, 1103.5501, 1610.00854, 1701.06102, 2105.12692}, or calibrate the relation at lower redshifts by other probes such as SN or CC in a model-independent way. Methods for such calibration include interpolation \cite{0802.4262}, cosmography  \cite{1003.5319, 1003.5755}, local regression \cite{1609.09631, 1610.00854}, Pad\'e approximation  \cite{1410.3960}, B\'ezier parametric curve \cite{1811.08934, 2010.05218, 2402.18967, 2408.02536, 2501.15233}, Gaussian process  \cite{2211.02473, 2212.14291, 2302.02559, 2307.16467, 2405.14357}, and other machine learning approaches \cite{2312.09440, 2502.10037, 2512.06454}.

We use the GRB data compiled by  \cite{2105.12692}, denoted as A118, covering the redshift range $0.33<z\leq 8.2$. This data set contains 118 long GRB, including 25 Fermi-GBM/LAT events
from  \cite{1910.07009}, and 93 events originally from \cite{1509.08558} and subsequently updated in \cite{1910.07009}. As noted in  \cite{2105.12692}, A118 is consistent with BAO and CC data used in that work.

The log likelihood is given by the following $\chi^2$ \cite{physics/0511182}
\be 
\chi^2_\text{GRB}= \sum_i^{N}\left(\frac{[y_i-y(x_i,z_i;a,b,\Theta)]^2}{\sigma^2_i}+\ln\sigma_i^2\right),
\ee
where  $\Theta$ stands for model parameters, which in this work include the PAge parameters, and
\be 
\sigma^2_i=\sigma_\text{int}^2+\sigma_{y,i}^2+b^2\sigma_{x,i}^2~,
\ee
with $\sigma_\text{int}$ denoting the intrinsic scatter of the Amati relation, treated as an additional parameter alongside $a$ and $b$, and the observational uncertainties given by
\be 
\sigma_{y,i}=\frac{\sigma_{E_\text{iso}}}{E_\text{iso}\ln 10 },\ \ \ \sigma_{x,i}=\frac{\sigma_{E_\text{p,i}}}{E_\text{p,i}\ln 10 }.
\ee
	
\section{Method I}
\label{sec:Method I}

We consider the following parametrizations of the CDDR violation, 
\ba 
\text{P1}: & \eta(z)&=1+\eta_0z,\\ 
\text{P2}: & \eta(z)&=1+\eta_0\frac{z}{1+z},\\
\text{P3}: & \eta(z)&=1+\eta_0\ln(1+z),\\ 
\text{P4}: & \eta(z)&=(1+z)^{\eta_0}.
\ea
In all the parametrizations above, $\eta_0<0$, $\eta_0=0$, and $\eta_0>0$ correspond to $\eta<1$, $\eta=1$, and $\eta>1$, respectively. 
P1 is the common limit of all these parametrizations for small $z$. Moreover, for $|\eta_0|<1$, the linear $\eta(z)$ deviates from 1 most rapidly with increasing $z$ among all these parametrizations. 
It is therefore expected that P1 is more sensitive to high-redshift data than the other forms of $\eta(z)$. In P2, the quantity $z/(1+z)\equiv y$ is sometimes called the $y$-redshift, which was introduced to improve the convergence at high redshifts in cosmography \cite{0710.1887,1104.3096}. For all $z\in(0,\infty)$, we have $y\in(0,1)$, thus the deviation from $1$ is less sensitive to high-redshift data. For small $|\eta_0|$, the logarithmic P3 and the power law P4 are almost equivalent, as will become apparent in the numerical results presented below.

A $z$-independent case $\eta(z)=1+\eta_0$ can also be considered (see, e.g.,  \cite{2504.10464}). However, this effectively corresponds to a recalibration of $M_B$,\footnote{
	This corresponds to a recalibration of $M_B$ when $\eta(z)$ enters through $d_L$ via $d_L=\eta(z)(1+z)d_M$, as in Eq. \eq{dL with eta}. In fact, $\eta(z)$ can also enter through $d_A$ as $d_A=d_M/(1+z)/\eta(z)$, with $d_L=(1+z)d_M$ unchanged. In that case, a constant $\eta(z)$ corresponds to a recalibration of the BAO scale $r_d$. 
} and is therefore not considered here. Further discussion can be found in Appendix \ref{app:const eta}.

The likelihood for different data combinations is constructed by combining the relevant $\chi^2$ defined above, as $\ln\mathcal{L}\propto-\sum_i\chi^2_i/2$, up to an inconsequential additive constant. The Markov Chain Monte Carlo (MCMC) analysis is performed using the package \texttt{emcee}\footnote{\url{https://github.com/dfm/emcee}}  \cite{emcee}. The posterior distributions are analyzed using the package \texttt{GetDist}\footnote{\url{https://github.com/cmbant/getdist}}  \cite{1910.13970}. Flat priors are assumed for the parameters as follows: $a\in(50,55)$, $b\in(0.8,1.4)$, $\sigma_\text{int}\in(0.3,0.6)$, $M_B\in(-20,-19)$, $r_d\in(130,160)$ Mpc, $p_{age}\in(0.2,1.2)$, and $\eta\in(-1,1)$, $H_0\in(60,80)$ km/s/Mpc. The range of the flat prior on $\eta_0$ depends on the chosen form of $\eta(z)$ and the maximum redshift of the data. It is selected to ensure that the parameter space is sufficiently sampled, while keeping $\eta(z)$ positive and not too far from unity. For example, when the GRB A118 data set is used, we choose $\eta_0\in(-0.1,0.1)$ for the linear P1, $\eta_0\in(-0.9,0.9)$ for the $y$-redshift P2, and $\eta_0\in(-0.4,0.4)$ for the logarithmic P3 and power-law P4. When GRB data are not included, the range can be correspondingly enlarged. Besides flat priors, Gaussian priors for $M_B$ and $r_d$ are also imposed over the same ranges, when necessary.

\subsection{Comparison of constraints with and without GRB data}

The GRB A118 data set is expected to place constraint at redshifts higher than those of the SN, BAO and CC data. To this end, we combine SN+BAO+CC+GRB, using PantheonPlus and DES Dovekie as the SN data sets, respectively. $M_B$ and $r_d$ are left as free parameters with flat priors, as previously discussed.

Since we constrain the uncalibrated GRB data together with cosmological parameters, we first present the results of the Amati relation parameters for different $\eta(z)$ in Table \ref{tab: Amati parameters}, where $a$, $b$, and $\sigma_\text{int}$ show negligible dependence on the choice of $\eta(z)$ or SN data set, indicating that the Amati relation is robust and the A118 data set is well standardized and self-consistent.

The cosmological parameters are listed in Table \ref{tab: SN BAO CC GRB}, which shows that adding the GRB data does not significantly improve the constraints.
This is because, unlike SN and BAO, the GRB  data typically have a large intrinsic scatter $\sigma_\text{int}\sim0.4$, which may reflect possible systematic uncertainties that are less well understood and modeled compared to SN and BAO \cite{2105.12692}. Consequently, the constraints from GRB are much weaker than those from the combination of the other probes. To illustrate this point, Figure \ref{fig: A118 comparison} shows the results for the linear $\eta(z)$ of P1 obtained from SN+BAO+CC and from A118 alone.\footnote{Note that applying the GRB data alone is not meaningful for testing the CDDR, since no $d_A$ information is available. The result of A118 alone is presented solely to illustrate its constraining power compared to SN+BAO+CC.}
The constraints from A118 alone on the PAge parameter $p_{age}$, $\eta$, $H_0$, and $\eta_0$ are considerably weaker than those obtained from the combined data sets. This is consistent with the observation made in  \cite{2105.12692}, where the GRB constraints are much weaker than those from BAO+CC. 
We have checked that for the other forms of $\eta(z)$, and when using DES Dovekie instead of PantheonPlus, the results are similar and are therefore not presented here. Moreover, the large intrinsic scatter is not unique to the data set A118. A different data set, A123, compiled in \cite{2405.14357}, is used to exhibit the same behavior, as discussed in Appendix \ref{app:A123}.  As a consequence, in the method adopted here, the uncalibrated GRB data provide very weak constraints compared to data from other well-established probes such as SN and BAO. Accordingly, the GRB data are not considered in the following.

	\begin{figure*}[tbp]
		\centering
		\includegraphics[scale=0.8]{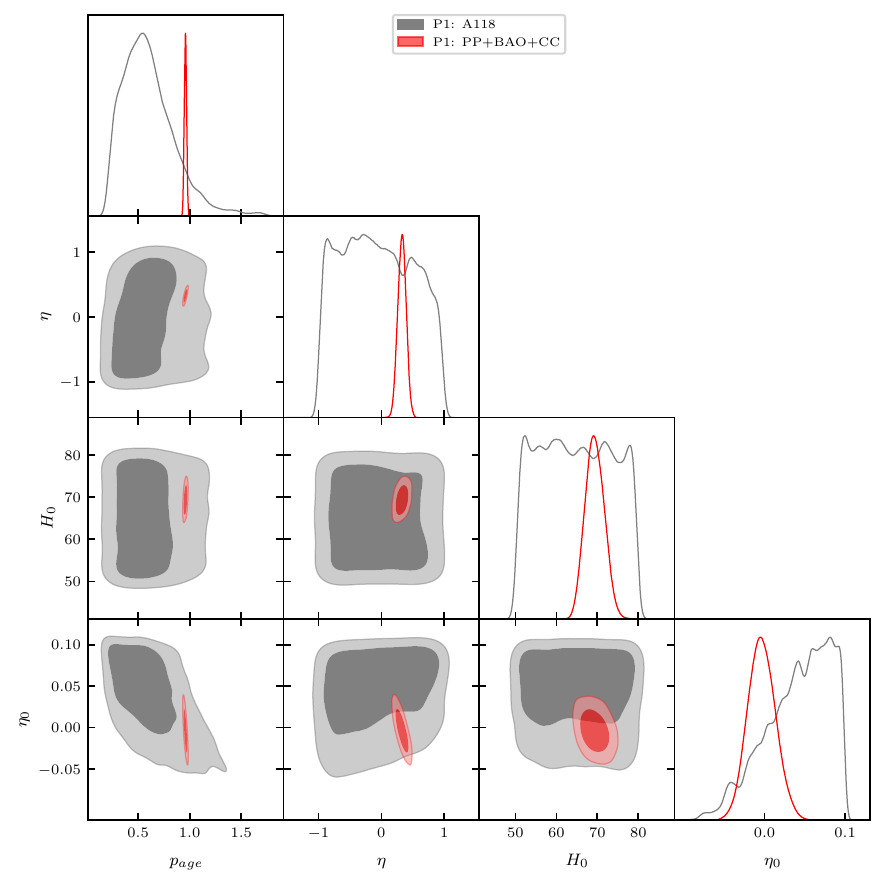} 
		\caption{Illustration of the weak constraints from the GRB data set A118 (grey) compared to PantheonPlus+BAO+CC (red), in the linear $\eta(z)$ case of P1. The combined results from PantheonPlus+BAO+CC+A118 are essentially the same as the red plots, and therefore are not shown for the sake of visual clarity.  }
		\label{fig: A118 comparison}
	\end{figure*}

In all cases, the mean values of $\eta_0$ are a bit below 0, indicating a tendency to $\eta<1$. Among other possibilities, this might imply the $d_L$ measurements are a bit smaller than the CDDR expectation, and the SNe appear brighter, if $d_A$ are measured accurately. Or, the $d_A$ measurements are a bit larger than the CDDR expectation, and the BAO length scales appear smaller, if $d_L$ are measured accurately. But, of course, no decisive conclusion can be reached so far, and the results are  still consistent with the CDDR within $1\sigma$. More stringent test involving more high quality data are worthy of further investigation.

	\begin{table*}[tbp]
		\renewcommand{\arraystretch}{1.5}
		\centering
		\caption{Constraints on the Amati relation parameters for different forms of $\eta(z)$. The data sets D$_\text{com}^\text{PP}$=PantheonPlus SN+BAO+CC, and D$_\text{com}^\text{DES}$=DES Dovekie SN+BAO+CC.}
		\label{tab: Amati parameters}
		\vspace{3mm}
	    \scalebox{0.85}{
			\begin{tabular}{ccccc}
				 
				\hline
				$\eta(z)$ & data set & $a$ & $b$ & $\sigma_\text{int}$\\
				\midrule[0.4pt]
				\multirow{2}{*}{P1} 
				& D$_\text{com}^\text{PP}$+A118 & $ 52.998 \pm 0.060$ & $ 1.133 \pm 0.087$ & $0.413^{+0.026}_{-0.031}$ \\  
				& D$_\text{com}^\text{DES}$+A118 & $ 53.000 \pm 0.061$ & $1.128 \pm 0.083$ & $0.412^{+0.025}_{-0.031}$ \\ 
				\midrule[0.4pt]
				\multirow{2}{*}{P2}
				& D$_\text{com}^\text{PP}$+A118 & $ 52.993 \pm 0.057$ & $1.128 \pm 0.084$ & $0.412^{+0.026}_{-0.032}$ \\  
				& D$_\text{com}^\text{DES}$+A118 & $52.994 \pm 0.057$ & $1.127 \pm 0.083$ & $0.412^{+0.026}_{-0.031}$ \\ 
				\midrule[0.4pt]
				\multirow{2}{*}{P3} 
				& D$_\text{com}^\text{PP}$+A118 & $ 52.990 \pm 0.059$ & $1.131 \pm 0.084$ & $0.412^{+0.026}_{-0.032}$ \\ 
				&D$_\text{com}^\text{DES}$+A118 & $ 52.991 \pm 0.058$ & $1.130 \pm 0.085$ & $0.412^{+0.026}_{-0.031}$ \\ 
				\midrule[0.4pt]
				\multirow{2}{*}{P4}
				& D$_\text{com}^\text{PP}$+A118 & $ 52.991 \pm 0.058$ & $1.131 \pm 0.084$ & $0.412^{+0.027}_{-0.031}$ \\  
				& D$_\text{com}^\text{DES}$+A118 & $ 52.991 \pm 0.058$ & $1.131 \pm 0.084$ & $0.412^{+0.027}_{-0.031}$ \\
				\hline
			\end{tabular}
		}
	\end{table*}
	
	\begin{table*}[tbp]
		\setlength\tabcolsep{3pt}
		\renewcommand{\arraystretch}{1.5}
		\centering
		\caption{Constraints on cosmological parameters, with and without A118. The common data sets D$_\text{com}^\text{PP}$=PantheonPlus SN+BAO+CC, and D$_\text{com}^\text{DES}$=DES Dovekie SN+BAO+CC.  $H_0$ and $r_d$ are in units of $\mathrm{km~s^{-1}~Mpc^{-1}}$ and $\mathrm{Mpc}$, respectively.}
		\label{tab: SN BAO CC GRB}
		\vspace{3mm}
		\scalebox{0.85}{
			\begin{tabular}{llccccccccc}
				
				\hline	
				
				$\eta(z)$ & data set & \(p_{\mathrm{age}}\)  & $\eta$ & $r_d$ &  $M_B$ & $H_0$  & $\eta_0$ \\
				\midrule[0.4pt]
				\multirow{4}{*}{P1} 
				& D$^\text{PP}_\text{com}$ & $ 0.958 \pm 0.011$ & $ 0.328 \pm 0.066$ & $ 144.5 \pm 4.4$ & $-19.373 \pm 0.071$ & $ 69.4 \pm 2.3$ & $-0.004 \pm 0.017$\\
				& D$^\text{PP}_\text{com}$+A118 & $0.958 \pm 0.010$ & $0.323 \pm 0.062$ & $144.9_{-5.0}^{+4.3}$ & $-19.380_{-0.070}^{+0.078}$ & $69.1 \pm 2.4$ & $-0.002 \pm 0.015$\\	
				
				&D$^\text{DES}_\text{com}$  & $0.958 \pm 0.012$ & $ 0.325 \pm 0.070$ & $144.6 \pm 4.5$ & $-19.342 \pm 0.072$ & $ 69.3^{+2.2}_{-2.5}$ & $-0.003 \pm 0.017$ \\
				
				&D$^\text{DES}_\text{com}$+A118 & $0.957 \pm 0.011$ & $ 0.320 \pm 0.065$ & $ 144.9 \pm 4.9$ & $-19.348 \pm 0.078$ & $ 69.1 \pm 2.5$ & $-0.002 \pm 0.016$ \\

				\midrule[0.4pt]
				\multirow{4}{*}{P2} 
				& D$^\text{PP}_\text{com}$ & $ 0.961 \pm 0.013$ & $  0.344 \pm 0.080$ & $ 144.9_{-5.1}^{+4.5}$ & $-19.369 \pm 0.079$ & $ 69.4 \pm 2.5$ & $-0.016 \pm 0.037$\\
				& D$^\text{PP}_\text{com}$+A118 & $0.961 \pm 0.012$ & $0.346 \pm 0.079$ & $144.9_{-5.0}^{+4.4}$ & $-19.369 \pm 0.078$ & $69.4 \pm 2.4$ & $-0.018 \pm 0.037$\\	
				
				&D$^\text{DES}_\text{com}$  & $ 0.962 \pm 0.013$ & $0.347 \pm 0.080$ & $144.8^{+4.5}_{-5.1}$  & $-19.332 \pm 0.081$ & $ 69.6 \pm 2.6$ & $-0.017 \pm 0.036$ \\
				
				&D$^\text{DES}_\text{com}$+A118& $0.962 \pm 0.013$ & $0.350 \pm 0.078$ & $ 144.8^{+4.5}_{-5.3}$ & $-19.332 \pm 0.081$ & $69.6 \pm 2.6$ & $-0.018 \pm 0.035$ \\

				\midrule[0.4pt]
				\multirow{4}{*}{P3} 
				& D$^\text{PP}_\text{com}$ & $ 0.959 \pm 0.012$ & $  0.335 \pm 0.074$ & $144.6 \pm 4.5$ & $-19.370 \pm 0.075$ & $69.4 \pm 2.4$ & $-0.009 \pm 0.027$\\
				& D$^\text{PP}_\text{com}$+A118 & $0.960 \pm 0.012$ & $0.342 \pm 0.073$ & $144.5\pm 4.9$ & $-19.366 \pm 0.079$ & $69.6 \pm 2.4$ & $-0.011 \pm 0.027$\\	
				&D$^\text{DES}_\text{com}$ & $0.960 \pm 0.012$ & $0.334 \pm 0.077$ & $ 144.7_{-5.0}^{+4.4}$ & $-19.338 \pm 0.078$ & $ 69.4 \pm 2.5$ &  $-0.007 \pm 0.026$\\
				
				&D$^\text{DES}_\text{com}$+A118  & $0.960 \pm 0.012$ & $0.336 \pm 0.074$ & $  144.6 \pm 4.8$ & $-19.335 \pm 0.079$ & $69.5 \pm 2.5$ & $-0.009 \pm 0.025$ \\
				
				\midrule[0.4pt]
				\multirow{4}{*}{P4} 
				& D$^\text{PP}_\text{com}$ & $ 0.960 \pm 0.012$ & $0.338 \pm 0.073$ & $144.7^{+4.6}_{-5.2}$ & $-19.370 \pm 0.080$ & $69.5 \pm 2.6$ & $-0.010 \pm 0.027$\\
				
				& D$^\text{PP}_\text{com}$+A118 & $0.960 \pm 0.012$ & $0.341 \pm 0.074$ & $144.8\pm 4.8$ & $-19.369 \pm 0.078$ & $69.5 \pm 2.5$ & $-0.012 \pm 0.026$\\
				
				&D$^\text{DES}_\text{com}$  & $0.960 \pm 0.012$ & $0.338 \pm 0.073$ & $144.7^{+4.6}_{-5.2}$  & $-19.370 \pm 0.080$ & $69.5 \pm 2.6$ & $-0.010 \pm 0.027$ \\
				
				&D$^\text{DES}_\text{com}$+A118 & $0.960 \pm 0.012$ & $0.341 \pm 0.074$ & $ 144.8 \pm 4.8$ & $-19.369 \pm 0.078$ & $69.5 \pm 2.5$ & $-0.012 \pm 0.026$ \\
				
				\hline	
			\end{tabular}
		}
	\end{table*}
	

Note the nearly identical results in the logarithmic P3 and power law P4, as shown in Table \ref{tab: SN BAO CC GRB}.  As is previously noted, since the CDDR violation is very small (with $|\eta_0|\sim 0.01$), it is easy to check that the difference of $\eta(z)$ in P3 and P4 is indeed negligible 
all the way up to $z\sim 10$. Therefore, only the logarithmic $\eta(z)$ of P3 will be considered in the following.

\subsection{Comparison of distance ladder and inverse distance ladder calibrations}

SN data provide measurements of relative distances, with $M_B$ to be calibrated and $H_0$ being degenerate with $M_B$. Similarly, BAO data provide distances relative to $r_d$, which is also degenerate with $H_0$. In this subsection, we consider several calibrations by combining the SN+BAO data with an $M_B$ prior from SH0ES and an $r_d$ prior from Planck, as well as the CC data. 
 
In the first combination, to calibrate the SN data and break the degeneracy, we use $M_B=-19.253\pm0.027$ from SH0ES \cite{2112.04510}, which was obtained using parallax measurements of nearby Cepheid variables. This is the common practice in the  distance ladder method. We also consider the combination following the inverse distance ladder method \cite{1811.02376, 2406.05049}. That is, we calibrate the SN and BAO data by imposing the Planck 2018  \cite{1807.06209} CMB (TT, TE, EE+lowE) result $r_d=147.05\pm 0.30$ Mpc \footnote{Following \cite{1811.02376}, CMB lensing is not involved in order to minimize the dependence on physics of the late-time universe.} 
as a Gaussian prior. Unlike the SH0ES result above, this $r_d$ result is obtained in $\Lambda$CDM, which may compromise the model independence of our results to some extent. In contrast, the CC data in the third combination effectively provide a length scale, and are independent of any particular cosmological model. Thus it can be used to break the degeneracy and calibrate the distances as a model-independent alternative to the $M_B$ prior and $r_d$ prior.

The results of the three calibrations are presented in Table~\ref{tab: calibrations PP} using PantheonPlus as the SN data set, and in Table~\ref{tab: calibrations DES} using DES Dovekie. The two tables show that, for all $\eta(z)$ parametrizations, the SH0ES $M_B$ prior yields the smallest $r_d$ value ($\sim 137$ and $139$ for PantheonPlus and DES Dovekie, respectively),\footnote{In the rest of this paragraph, whenever two values are given in parentheses, they correspond to PantheonPlus and DES Dovekie, respectively.} and the largest values of $M_B$ ($\sim -19.25$) and $H_0$ ($\sim 73$ and $72$). In contrast, the Planck $r_d$ prior yields the largest $r_d$ value ($\sim 147$), and the smallest values of $M_B$ ($\sim -19.40$ and $-19.37$) and $H_0$ ($\sim 68$), exhibiting a tension at the $\sim5\sigma$ level. Combining the CC data yields intermediate values, albeit with larger uncertainties, roughly bridging the other two results. Note that the precision of the current CC data  is lower than that of the SN, BAO, and CMB data, therefore the calibration using CC leads to larger uncertainties on $r_d$, $M_B$, and $H_0$, as seen from the tables.

On the other hand, the three different calibrations above have no significant influence on the CDDR violation parameter $\eta_0$ (both mean value and uncertainty), or on the PAge parameter $p_{age}$ and $\eta$, for all parametrizations of $\eta(z)$. All $\eta_0$ results are consistent with no CDDR violation within $1\sigma$. However, this does not mean the CDDR is independent of all calibration choices. To illustrate this point, the results of simultaneously imposing the SH0ES $M_B$ and Planck $r_d$ as priors are listed in Table \ref{tab:eta0 MB rd}, where it is clear that the CDDR is violated at  the $3\sigma$ to $4\sigma$ level. 
This suggests that an apparent CDDR violation may also reflect the inconsistency between results from different probes, which, in this case, correspond to the early-time measurements from the CMB and the late-time measurements from SNe and Cepheids, as widely discussed in the context of the Hubble tension literature (e.g., \cite{2101.08641,2103.08723,2407.18292, 2504.01669}).\footnote{Note that an alternative perspective proposed in \cite{2504.10464} reverses the argument and interprets the CDDR violation as a phenomenological explanation of the Hubble tension. In this interesting picture, the CDDR is violated in such a way that the SH0ES $M_B$ and Planck $r_d$ can be reconciled in the $\Lambda$CDM model.} Similar effects have also been identified in the impact of different $r_d$ and $M_B$ values on the CDDR test using PantheonPlus+DESI DR2 \cite{2507.11518}, and on the consistency test between SN and BAO data using DES5yr/Dovekie and DESI DR2 \cite{2510.04179}.

	\begin{table*}[tbp]
		\setlength\tabcolsep{3pt}
		\renewcommand{\arraystretch}{1.5}
		\centering
		\caption{Constraints from different calibrations. D$_\text{com}$=PantheonPlus SN+BAO. Planck $r_d=147.05 \pm 0.30$. SH0ES $M_B=-19.253\pm0.027$. $H_0$ and $r_d$ are in units of $\mathrm{km~s^{-1}~Mpc^{-1}}$ and $\mathrm{Mpc}$, respectively.}
		\label{tab: calibrations PP}
		\vspace{3mm}
		\scalebox{0.85}{
			\begin{tabular}{llccccccccc}
				\hline
				
				$\eta(z)$ & data set & \(p_{\mathrm{age}}\)  & $\eta$ & $r_d$ &  $M_B$ & $H_0$  & $\eta_0$ \\
				\midrule[0.4pt]
				\multirow{3}{*}{P1} 
				& D$_\text{com}$+$M_B$ & $0.958 \pm 0.011$ & $ 0.325 \pm 0.066$ & $136.8 \pm 2.4$ & $-19.254 \pm 0.027$ & $73.23 \pm 0.93$ & $-0.003 \pm 0.017$\\
				& D$_\text{com}$+CC & $0.958 \pm 0.011$ & $0.328 \pm 0.066$ & $144.5 \pm 4.4$ & $-19.373 \pm 0.071$ & $69.4 \pm 2.3$ & $-0.004 \pm 0.017$
				\\ 
				& D$_\text{com}$+$r_d$ & $  0.958 \pm 0.011$ & $0.326 \pm 0.066$ & $147.05 \pm 0.30$  & $-19.411 \pm 0.027$ & $ 68.11 \pm 0.93$ & $-0.003 \pm 0.017$ \\  
				
				\midrule[0.4pt]
				\multirow{3}{*}{P2}  
				& D$_\text{com}$+$M_B$ & $ 0.961 \pm 0.013$ & $ 0.345 \pm 0.079$ & $137.3 \pm 2.8$ & $-19.253 \pm 0.024$ & $ 73.21 \pm 0.94$ & $-0.016\pm 0.038$ \\
				& D$_\text{com}$+CC & $ 0.961 \pm 0.013$ & $0.344 \pm 0.080$ & $144.9^{+4.9}_{-5.1}$ & $-19.369 \pm 0.079$ & $69.4\pm 2.5$ & $-0.016 \pm 0.037$
				\\ 
				& D$_\text{com}$+$r_d$ & $   0.960 \pm 0.013$ & $ 0.338 \pm 0.080$ & $147.04 \pm 0.30$  & $-19.404 \pm 0.035$ & $  68.3 \pm 1.1$ & $-0.013\pm 0.039$ \\ 
				
				\midrule[0.4pt]
				\multirow{2}{*}{P3} 
				& D$_\text{com}$+$M_B$ & $0.960 \pm 0.012$ & $0.335 \pm 0.076$ & $137.0 \pm 2.6$ & $-19.253 \pm 0.027$ & $73.23 \pm 0.95$ & $-0.008 \pm 0.028$ \\
				& D$_\text{com}$+CC & $0.959 \pm 0.012$ & $0.335 \pm 0.074$ & $144.6 \pm 4.5$ & $-19.370 \pm 0.075$ & $ 69.4 \pm 2.4$ & $-0.009 \pm 0.027$
				\\ 
				& D$_\text{com}$+$r_d$ & $  0.960 \pm 0.012$ & $ 0.337 \pm 0.075$ & $147.05 \pm 0.30$  & $-19.406 \pm 0.033$ & $ 68.3 \pm 1.0$ & $-0.009 \pm 0.027$ \\

				\hline
			\end{tabular}
		}
	\end{table*}
	

	\begin{table*}[tbp]
		\setlength\tabcolsep{3pt}
		\renewcommand{\arraystretch}{1.5}
		\centering
		\caption{Constraints from different calibrations. The same as Table \ref{tab: calibrations PP}, except for replacing PantheonPlus by DES Dovekie, i.e. D$_\text{com}$=DES Dovekie SN+BAO. }
		\vspace{3mm}
		\label{tab: calibrations DES}
		\scalebox{0.85}{
			\begin{tabular}{llccccccccc}
				\hline
				
				$\eta(z)$ & data set & \(p_{\mathrm{age}}\)  & $\eta$ & $r_d$ &  $M_B$ & $H_0$  & $\eta_0$ \\
				\midrule[0.4pt]
				\multirow{3}{*}{P1} 
				& D$_\text{com}$+$M_B$ & $ 0.958 \pm 0.012$ & $ 0.325 \pm 0.070$ & $138.8 \pm 2.5$ & $-19.253 \pm 0.026$ & $72.15 \pm 0.95$ & $-0.003 \pm 0.017$\\
				& D$_\text{com}$+CC  & $0.958 \pm 0.012$ & $ 0.325 \pm 0.070$ & $144.6 \pm 4.5$ & $-19.342 \pm 0.072$ & $69.3_{-2.5}^{+2.2}$ & $-0.003 \pm 0.017$
				\\ 
				& D$_\text{com}$+$r_d$ & $ 0.958 \pm 0.012$ & $ 0.324 \pm 0.073$ & $ 147.05 \pm 0.31$ & $-19.378 \pm 0.029$ & $68.1 \pm 1.0$ & $-0.003 \pm 0.017$ \\

				\midrule[0.4pt]
				\multirow{3}{*}{P2}  
				& D$_\text{com}$+$M_B$ & $0.961 \pm 0.013$ & $ 0.346\pm 0.081$ & $139.5 \pm 2.9$ & $-19.253 \pm 0.027$ & $72.11 \pm 0.97$ & $-0.016 \pm 0.036$ \\
				& D$_\text{com}$+CC  & $ 0.962 \pm 0.013$ & $0.347 \pm 0.080$ & $144.8^{+4.5}_{-5.1}$  & $-19.332 \pm 0.081$ & $ 69.6 \pm 2.6$ & $-0.017 \pm 0.036$ \\
				& D$_\text{com}$+$r_d$ & $0.962 \pm 0.013$ & $0.345 \pm 0.082$ & $147.06 \pm 0.30$  & $-19.368 \pm 0.037$ & $68.4 \pm 1.1$ & $-0.015 \pm 0.037$ \\

				\midrule[0.4pt]
				\multirow{3}{*}{P3} 
				& D$_\text{com}$+$M_B$ & $0.959 \pm 0.013$ & $0.330 \pm 0.079$ & $ 139.0 \pm 2.7$ & $-19.253 \pm 0.026$ &  $72.15 \pm 0.95$ & $-0.006 \pm 0.027$ \\
				& D$_\text{com}$+CC & $0.960 \pm 0.012$ & $0.334 \pm 0.077$ & $ 144.7_{-5.0}^{+4.4}$ & $-19.338 \pm 0.078$ & $ 69.4 \pm 2.5$ &  $-0.007 \pm 0.026$
				\\ 
				& D$_\text{com}$+$r_d$ & $ 0.960 \pm 0.013$ & $0.332 \pm 0.079$ & $147.05 \pm 0.30$ & $-19.375 \pm 0.034$ & $68.2 \pm 1.1$  & $-0.007 \pm 0.026$\\

				\hline
			\end{tabular}
		}
	\end{table*}
	
	
	\begin{table*}[tbp]
		\setlength\tabcolsep{3pt}
		\renewcommand{\arraystretch}{1.5}
		\centering
		\caption{$\eta_0$ for SH0ES $M_B$+Planck $r_d$ priors. The data sets are SN+BAO, with SN being PantheonPlus (left column) and DES Dovekie (right column), respectively. The deviations from $\eta_0=0$ are around $3\sigma\sim 4\sigma$. }
		\vspace{3mm}
		\label{tab:eta0 MB rd}
		\scalebox{0.85}{
			\begin{tabular}{lccccccccc}
				\hline
				
				$\eta(z)$   & $\eta_0$ (PP) & $\eta_0$ (DES)  \\
				\midrule[0.4pt]
				 P1 & $-0.045 \pm 0.013$ &  $-0.039 \pm 0.013$ \\
				
				\midrule[0.4pt]
				 P2 & $-0.104 \pm 0.024$  & $-0.086 \pm 0.023$ \\
				
				\midrule[0.4pt]
				 P3 & $-0.072 \pm 0.019$ & $-0.059 \pm 0.017$   \\

				\hline
			\end{tabular}
		}
	\end{table*}
	
	\subsection{Comparison between PantheonPlus and DES Dovekie}
	 
	To compare the two SN data sets, PantheonPlus and DES Dovekie, the posterior distributions of the SN+BAO+CC results in Tables \ref{tab: calibrations PP} and \ref{tab: calibrations DES} are shown in Figures \ref{fig:SN compare linear}, \ref{fig:SN compare y}, and \ref{fig:SN compare log}. The results for the other two combinations, SN+BAO+$M_B$ and SN+BAO+$r_d$, are similar and thus are not presented. All these figures exhibit almost identical features. In particular, for the CDDR violation parameter $\eta_0$, as well as many other parameters, the two SN data sets produce essentially the same results. The only exception is the parameter $M_B$, for which significant differences can be observed in the 1D and 2D plots in the first column of each figure. In particular, the contours on the $M_B$-$H_0$ plane show features similar to those in \cite{2505.22369}, where the PAge model was constrained by PantheonPlus+DESI DR2+CC and DES5yr+DESI DR2+CC (as well as other data combinations). \footnote{Since the CDDR violation is very small, as discussed above, the results here are comparable to those in \cite{2505.22369}, which correspond to the case with fixed $\eta_0=0$.} By comparing the figures presented here with Figure~1 in \cite{2505.22369}, one can see that the discrepancy between PantheonPlus and DES5yr is noticeably reduced when DES5yr is replaced by DES Dovekie. In other words, the latest DES Dovekie is more compatible with PantheonPlus than DES5yr. This is related to the so-called $a_B$ tension \cite{2103.08723,2401.14170,2410.06053}, and further discussion is deferred to Appendix \ref{app: aB}.
	
	\begin{figure*}[tbp]
		\centering
		\includegraphics[scale=0.8]{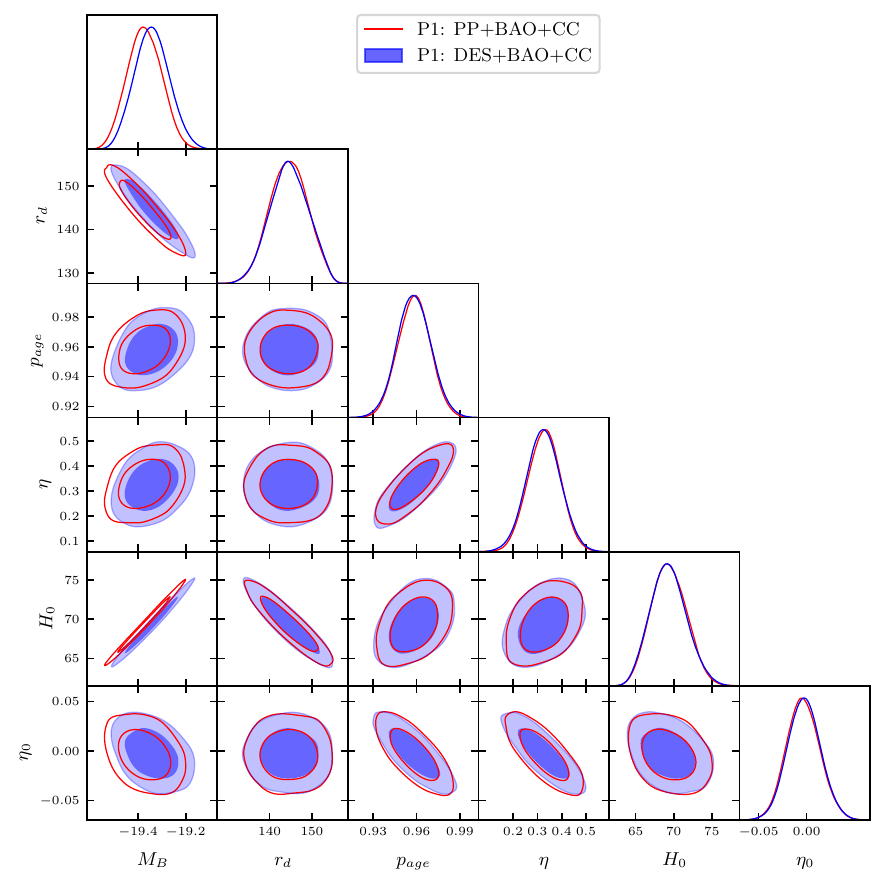} 
		\caption{Comparison of the constraints from PantheonPlus+BAO+CC (red) and DES Dovekie+BAO+CC (blue), in the linear $\eta(z)$ case of P1.  }
		\label{fig:SN compare linear}
	\end{figure*}
	
	\begin{figure*}[tbp]
		\centering
		\includegraphics[scale=0.8]{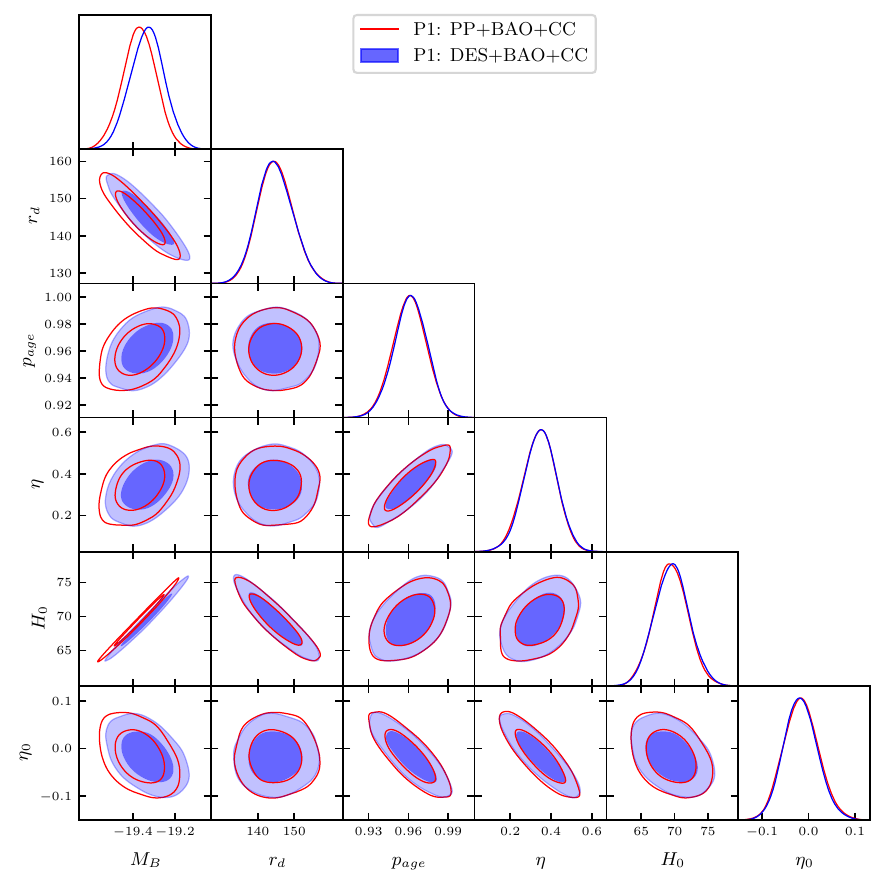} 
		\caption{Comparison of the constraints from PantheonPlus+BAO+CC (red) and DES Dovekie+BAO+CC (blue), in the $y$-redshift $\eta(z)$ case of P2.  }
		\label{fig:SN compare y}
	\end{figure*}
	
	\begin{figure*}[tbp]
		\centering
		\includegraphics[scale=0.8]{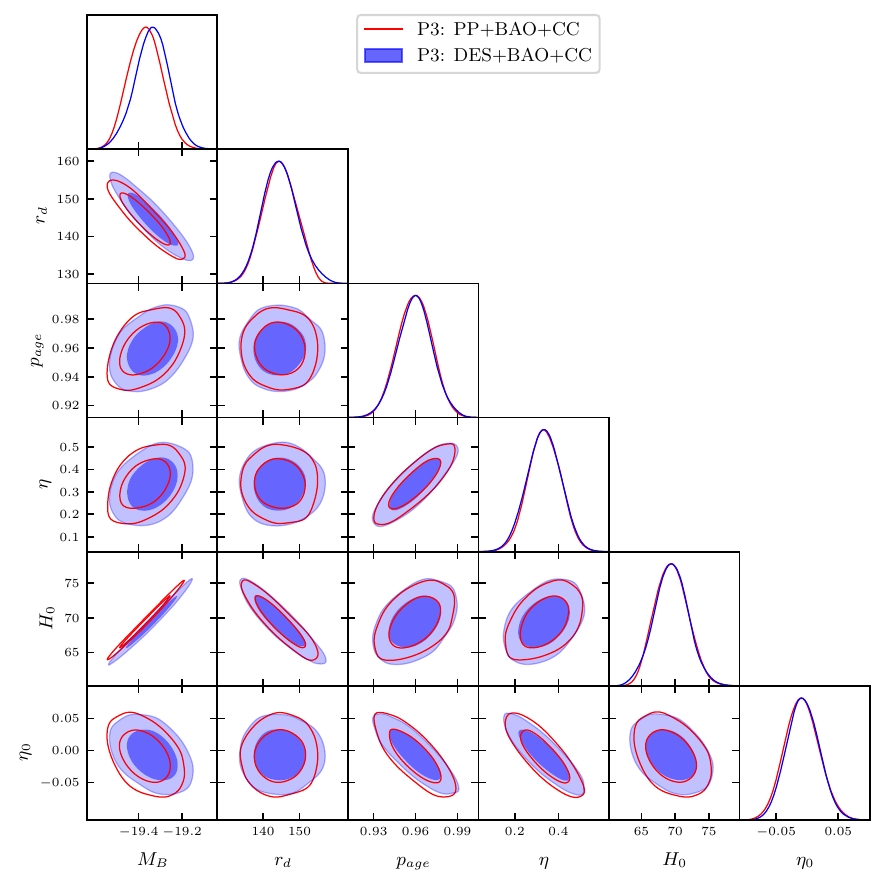} 
		\caption{Comparison of the constraints from PantheonPlus+BAO+CC (red) and DES Dovekie+BAO+CC (blue), in the logarithmic $\eta(z)$ case of P3.  }
		\label{fig:SN compare log}
	\end{figure*}
	
\section{Method II}
\label{sec:Method II}
 
In the previous section, with Method I, tight constraints can be obtained from observational data including SN, BAO, and CC. Although the GRB data have comparatively weaker constraining power, they can indeed be used in principle to provide constraints from higher redshifts. However, this method does require \textit{a priori} parametric forms of the CDDR violation $\eta(z)$, which may introduce some bias. To enable a non-parametric construction of the CDDR violation, one can resort to Method II, of which the basic strategy is as follows: (i) reconstruct $d_L$ from SN  data; (ii) construct $\eta_\text{obs}$ at the BAO redshifts $z_\text{BAO}$, with which one can then either perform non-parametric reconstruction of $\eta(z)$ using data-driven methods such as GP, or constrain given parametric forms of $\eta(z)$.

This method requires that the observational measurements leading to $d_A$ and $d_L$ be within the same redshift range.
Since no current cosmological probe provides $d_A$ data extending to the high redshifts probed by GRB, we do not use GRB data and focus only on SN and BAO data. Unlike PantheonPlus, the most recent DES Dovekie SN data set has not been widely used to study the CDDR. Therefore here we perform a CDDR test using DES Dovekie and, for comparison, PantheonPlus, following the method in  \cite{2407.12250,2506.17926,2508.07040,2507.13811}.

Unlike Method I in Section \ref{sec:Method I}, only the transverse BAO data $d_M/r_d$ can be used. Moreover, since we only reconstruct $d_L$ (or more precisely, $\tilde d_L$ defined below) within the range of the SN data, the BAO data points beyond this range cannot be used to construct $\eta_\text{obs}$. As a result, for PantheonPlus and DES Dovekie, only five and three data points are used, respectively, at $z_\text{BAO}=\{0.510, 0.706, 0.934, 1.321, 1.484\}$ and $z_\text{BAO}=\{0.510, 0.706, 0.934\}$.  To compare with the results obtained using Method I, we only use $\eta_\text{obs}$ to constrain the same P1, P2, and P3 parametrizations of $\eta(z)$ studied above, and leave the non-parametric reconstruction of $\eta(z)$  to future work with increased number of $d_A$ data.

In what follows, $d_L$ is reconstructed using the GP method.  We use the package \texttt{GaPP}\footnote{
	The original code written for Python~2.7 is available at \url{https://github.com/carlosandrepaes/GaPP}

	An updated version for Python~3 is available at \url{https://github.com/lighink/GaPP3}}
\cite{1204.2832}.
For details of the GP method, see \cite{1204.2832,1311.6678}. We adopt the widely used squared exponential also called the radial basis function) kernel. The two hyperparameters are optimized rather than marginalized over, as is commonly done in the literature.

Defining $\tilde d_M\equiv d_M(z)/r_d$ as the BAO observable, and $\tilde d_L(z)\equiv 10^{0.2 m_B-5}$ as the quantity to be reconstructed via GP, we can write the CDDR violation as
\be 
\eta(z)=\frac{\tilde d_L 10^{-0.2M_B}}{\tilde d_M(z) r_d (1+z)}.
\ee
As is widely discussed in the Hubble tension literature (e.g.,  \cite{2103.01183, 2106.15656, 2401.13187}), the values of the late-time $M_B$ and early-time $r_d$ are related to the inconsistency in the $H_0$ values. Therefore, one should avoid simultaneously setting $M_B$ to the SH0ES value and $r_d$ to the Planck value. To circumvent this issue, we define a dimensionless parameter $\zeta\equiv 10^{-0.2M_B}\text{Mpc}/r_d$ to characterize the degenerate parameters $M_B$ and $r_d$, and construct $\eta_\text{obs}$ from the observables as
\be 
\eta_\text{obs}(z,\zeta)=\frac{\tilde d_L(z)\zeta}{\tilde d_M(z) (1+z)},
\ee
where $\tilde d_M(z)$ is given by BAO measurements, and $\tilde d_L(z)$ is constructed using GP directly from the SNe data to provide values at the BAO redshifts. A similar discussion can be found in  \cite{2506.17926}. The difference here is that we use the latest DES Dovekie as well as PantheonPlus (see also \cite{2507.13811} where a similar problem was studied using an artificial neural network method instead of the GP method). The GP reconstructions of $\tilde d_L(z)$ from PantheonPlus and DES Dovekie are respectively presented in Figure \ref{fig:GP}. Note that we use the SN data to reconstruct $\tilde d_L$ instead of $m_B$, because the GP reconstruction of $m_B$ from PantheonPlus shows unphysical wiggles (recall that $m_B$ is expected to be a monotonic function of $z$). Further discussion on this point is deferred to Appendix \ref{app:GP for mB}, where we also show that for DES Dovekie, there are no significant wiggles in $m_B(z)$ and the result is essentially the same as the direct reconstruction of $\tilde d_L$ as presented in Figure \ref{fig:GP}. 
	
	\begin{figure*}[tbp]
		\centering
		\includegraphics[scale=0.33]{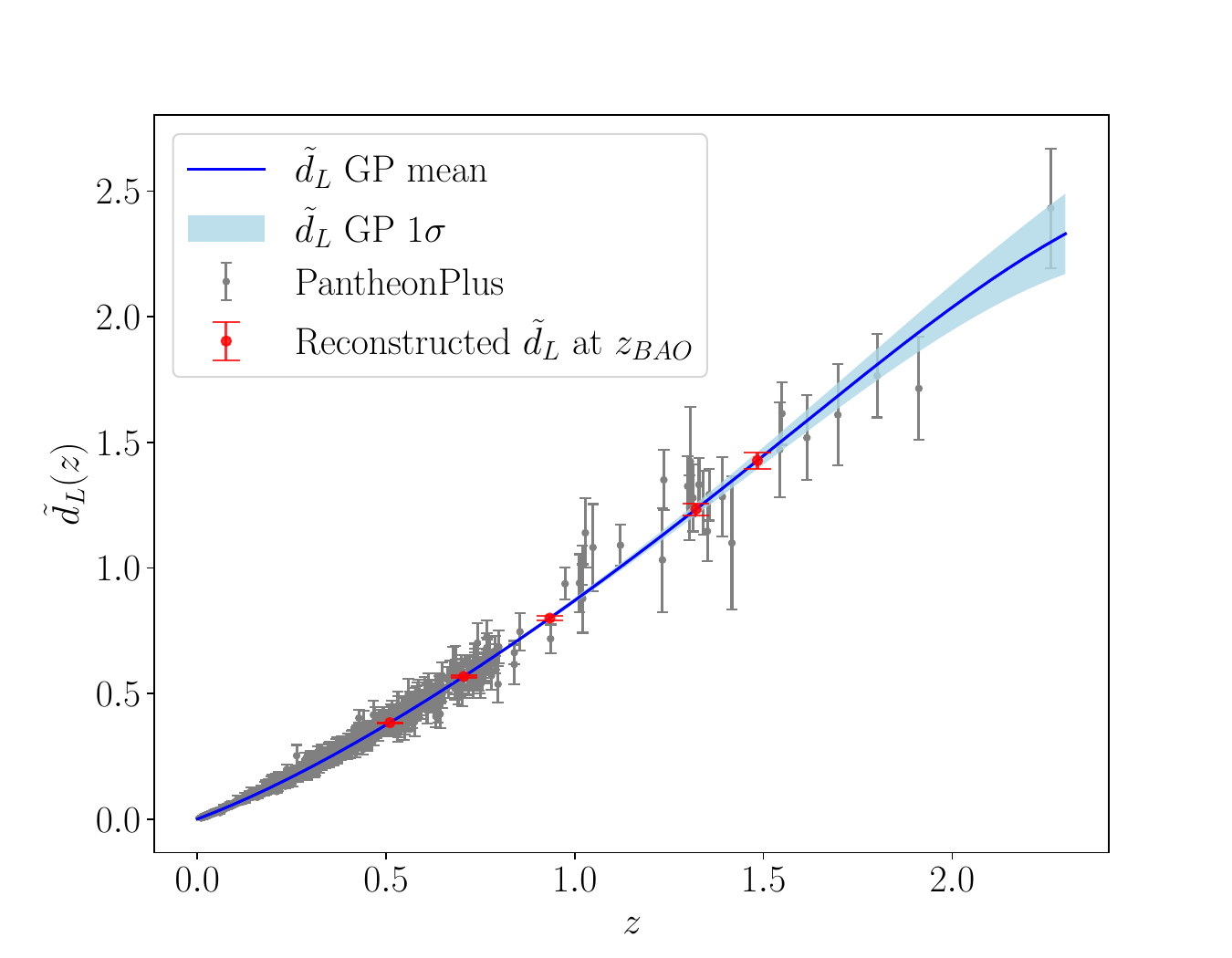} 
		\includegraphics[scale=0.33]{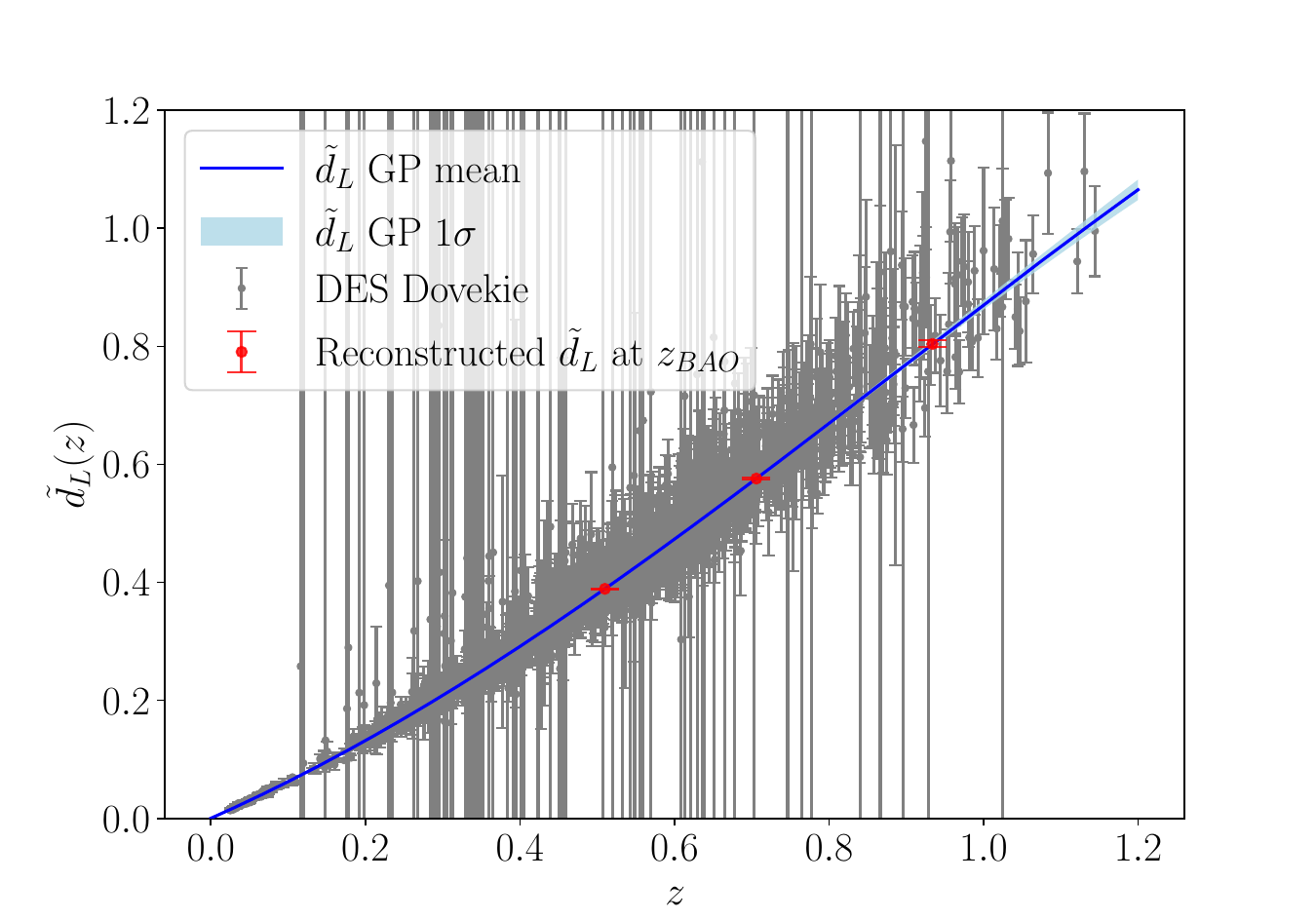}
		\caption{GP reconstruction of $\tilde d_L(z)$ from PantheonPlus (left) and DES Dovekie (right). The mean value and $1\sigma$ range are shown in blue. The reconstructed $\tilde d_L(z)$ values at the BAO redshifts are plotted in red. The SN data points are shown in grey. The extremely large error bars of some DES Dovekie data points in the right panel correspond to possible non-type Ia contaminants whose uncertainties are weighted by the BEAMS probability and thus get inflated, as is discussed in Section \ref{subsection:SN}.   }
		
		\label{fig:GP}
	\end{figure*}

Then one can easily construct five and three $\eta_\text{obs}$ values from PantheonPlus+BAO and DES Dovekie+BAO, respectively, and constrain the CDDR violation $\eta(z)$ via
\be 
\chi^2_\eta=\sum_i^{\text{BAO}}\left(\frac{[\eta(z_i,\eta_0)-\eta_\text{obs}(z_i,\zeta)]^2}{\sigma_\eta^2(z_i,\zeta)} +\ln\sigma_\eta^2\right),
\ee
where  $\zeta$ and $\eta_0$ are two parameters to be constrained, \footnote{
	Note that the nuisance parameter $\zeta$ can also be analytically marginalized over, as was done in \cite{2407.12250, 2506.17926,2508.07040}, following the usual practice in dealing with the nuisance parameters of SN data \cite{1104.1443}. We will not marginalize over $\zeta$ here, since we want to emphasize the influence of its value on the CDDR violation.} 
and $\sigma_{\eta}$ can be calculated from $\sigma_{\tilde d_L}$ and $\sigma_{\tilde d_M}$ by propagation of uncertainty, \footnote{Another interesting way to deal with these two nuisance parameters is to follow  \cite{2301.02997} (see also \cite{2405.12142, 2506.12759}) to construct a relative estimator $\eta_{ij}(z)=\eta(z_i)/\eta(z_j)$, where $r_d$ and $M_B$ are simply canceled out. We do not pursue this and leave it for future work.
}
\be 
\sigma_\eta^2=\left(\frac{\sigma_{\tilde d_L}\zeta}{\tilde d_M(1+z)}\right)^2+\left(\frac{\sigma_{\tilde d_M} \tilde d_L\zeta}{\tilde{d}_M^2(1+z)}\right)^2.
\ee

The results are listed in Table \ref{tab: eta0 zeta} and plotted in Figure \ref{fig:eta0 zeta}. 
Compared to the results from Method I, the uncertainties are larger, because of the flat prior on $\zeta$, and much smaller number of data points used in constraining $\eta_0$, i.e., the constructed five and three $\eta_\text{obs}$ values for PantheonPlus and DES Dovekie, respectively. This is also responsible for the larger uncertainties using DES+BAO than using PP+BAO as shown in Table \ref{tab: eta0 zeta}. An important lesson learned from these figures is that there is a correlation between $\eta_0$ and $\zeta$.

	\begin{table*}[tbp]
		\setlength\tabcolsep{3pt}
		\renewcommand{\arraystretch}{1.5}
		\centering
		\caption{Constraints obtained using Method II with PantheonPlus and DES Dovekie respectively combined with BAO, for P1, P2, and P3 parametrizations of $\eta(z)$. }
		\label{tab: eta0 zeta}
		\vspace{3mm}
		\scalebox{0.85}{
			\begin{tabular}{cccccccccc}
				
				\hline
				
				$\eta(z)$ & data set & $\eta_0$ & $\zeta$ \\
				\midrule[0.4pt]
				\multirow{2}{*}{P1} 
				& PP+BAO & $0.022 \pm 0.025$ & $53.5 \pm 1.1$ \\
				& DES+BAO & $0.035 \pm 0.040$ & $53.3 \pm 1.6$ \\ 
				
				\midrule[0.4pt]
				\multirow{2}{*}{P2} 
				& PP+BAO & $0.094 \pm 0.092$ & $54.7 \pm 2.1$ \\
				& DES+BAO & $0.121 \pm 0.116$ & $54.6 \pm 2.5$ \\

				\midrule[0.4pt]
				\multirow{2}{*}{P3} 
				& PP+BAO & $0.048 \pm 0.050$ & $54.0 \pm 1.6$ \\
				& DES+BAO & $0.066 \pm 0.070$ & $53.8 \pm 2.0$ \\ 
				
				
				\hline
			\end{tabular}
		}
	\end{table*}
	
	\begin{figure*}[tbp]
		\centering
		\includegraphics[scale=0.4]{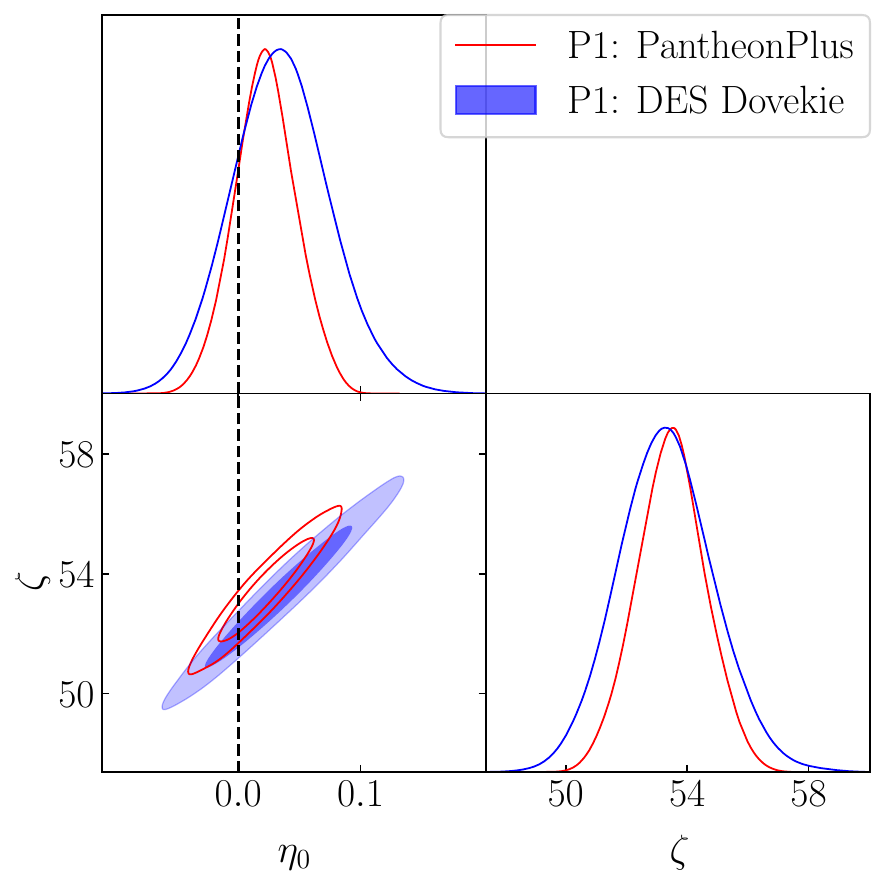} 
		\includegraphics[scale=0.4]{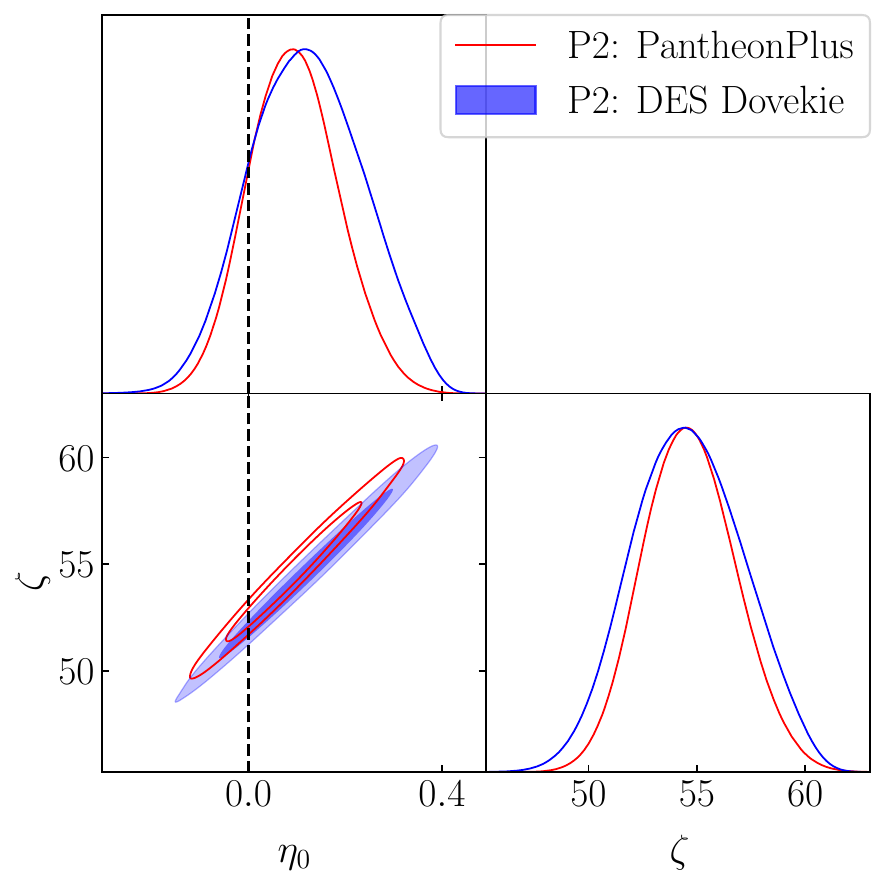} 
		\includegraphics[scale=0.4]{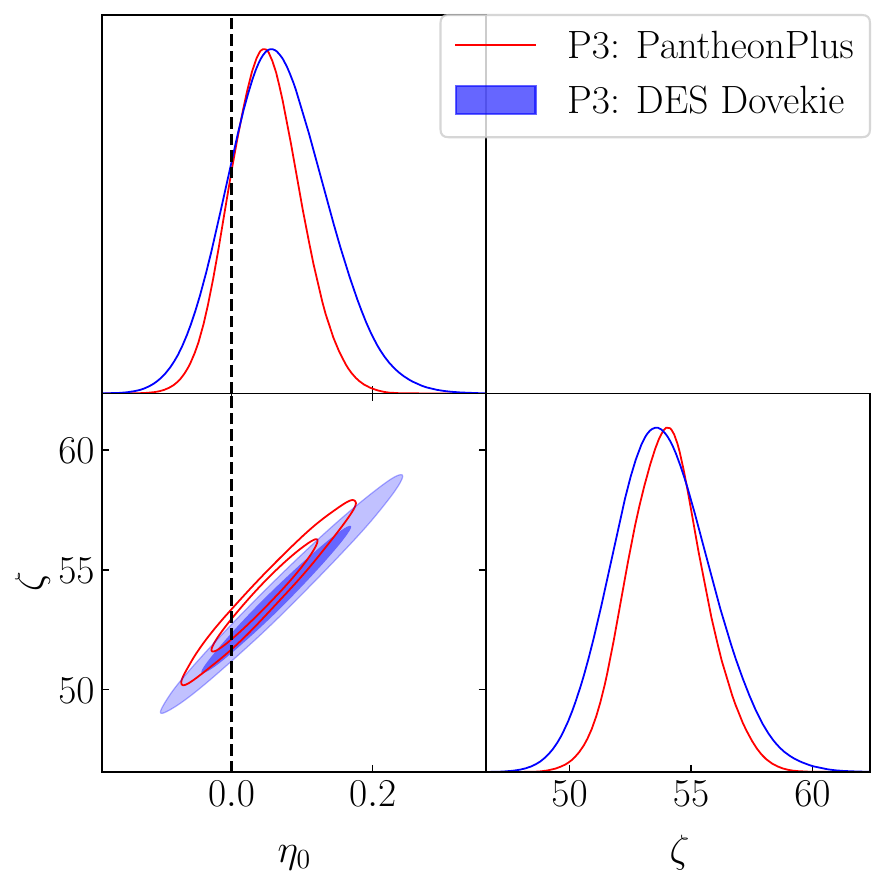} 

		\caption{Posterior distributions using Method II with PantheonPlus and DES Dovekie respectively combined with BAO, for parametrizations of $\eta(z)$ for P1, P2, and P3. The vertical dashed line corresponds to $\eta_0=0$ for the validity of the CDDR. }
		\label{fig:eta0 zeta}
	\end{figure*}

Compared to the results in Section \ref{sec:Method I}, the $\eta_0$ values here seem to suggest a strong tendency toward positive $\eta_0$. The difference arises because $\zeta$, which combines $M_B$ and $r_d$, is left as a free parameter to be fitted without any physical input at all. In contrast, $M_B$ and $r_d$ in Section \ref{sec:Method I} are constrained with physical input,  either from the SH0ES or Planck result as a prior, or from CC data. To conform with the previous section, one should consider imposing physical constraint on $\zeta$ (i.e., on the specific combination of $r_d$ and $M_B$). As mentioned before, the values of $r_d$ and $M_B$ represent the physics of the early-time and late-time universe, respectively, and they are closely related to the Hubble tension. Thus, once again, simultaneously imposing SH0ES $M_B$ and Planck $r_d$ leads to $\eta_\text{obs}$ exhibiting significant violation of the CDDR, as can be seen in Figure \ref{fig:eta errobar}, where the $\eta_\text{obs}$ values constructed from PantheonPlus (black) and DES Dovekie (grey) with these two priors exhibit deviations from $\eta_0=0$ at approximately the $3\sigma$ to $4\sigma$ level.

Many consistent combinations of $M_B$ and $r_d$ can be used. For example, one may use the Planck result for $r_d$, and $M_B=-19.363\pm 0.016$ obtained in \cite{2208.14740} by applying GP to SN+BAO data, as in \cite{2506.17926,2507.13811}. Since a community consensus on a satisfactory solution of the Hubble tension has not yet been reached, a pragmatic way adopted here is to use the values of $M_B$ and $r_d$ obtained in the previous section as listed in Table \ref{tab: calibrations PP} and \ref{tab: calibrations DES}.  As can be seen from the tables, for each $\eta(z)$ parametrization, the constraints on $\eta_0$ are essentially independent of the three calibrations. Moreover, the values of $r_d$ and $M_B$ obtained in different calibrations result in consistent values of $\zeta$. By imposing these $r_d$ and $M_B$ values to construct $\zeta$ and $\eta_\text{obs}$ via Monte Carlo sampling, and then constraining $\eta_0$ via MCMC, we obtain the constraints on $\eta_0$, which are presented in Table \ref{tab:eta0 physical zeta} and Figure \ref{fig:eta errobar}. The values of $\zeta$ are indeed all consistent within approximately $1\sigma$ across the three parametrizations of $\eta(z)$ and the two SN data sets. The mean values of $\zeta$ are approximately $51$ to $52$, which are smaller than the results without physical inputs ($53$ to $54$) listed in Table \ref{tab: eta0 zeta}. These values lie marginally within the $1\sigma$ range of the previous results, as also shown in Figure \ref{fig:eta0 zeta}.

Moreover, the physical inputs lead to negative mean values for $\eta_0$, similar to the results in Section \ref{sec:Method I}. For each parametrization of $\eta(z)$, the different calibrations yield consistent mean values of $\eta_0$, but with different uncertainties. The calibration using SN+BAO+CC yields the largest uncertainties, such that the CDDR is valid within $1\sigma$. The other two calibrations produce smaller uncertainties, indicating a marginal deviation from the CDDR at about $1\sigma$, while remaining consistent with the CDDR within $2\sigma$. Note, however, that the values of $\eta_0$ here do not exactly match those in the previous section. This is because the data used in the two methods are not exactly the same. Here, instead of the full BAO data used in the previous section, only five (for PantheonPlus) and three (for DES Dovekie) transverse BAO data points are used. More importantly, the data are used in different ways. In Method I, all data are used on the same footing to jointly constrain the parameters by MCMC. In contrast, in this section, the SN data are first used to reconstruct $\tilde d_L$ via GP, and then the reconstruction results are combined with the BAO data (with a reduced number of data points) to constrain the parameter $\eta_0$ via MCMC.

In addition, one can also see from Table \ref{tab:eta0 physical zeta} that, although the $\eta_0$ values are basically consistent between the two SN data sets, the mean values of $\eta_0$ obtained using DES Dovekie are roughly $1.5$ to $2$ times that obtained using PantheonPlus. 
This difference can be partly attributed to the smaller number of $\eta_\text{obs}$ constructed using DES Dovekie. Indeed, the absence of data points at the two BAO redshifts may have a significant impact. Another cause of the difference is related to the values of $\zeta$, which depend on $M_B$ and $r_d$ obtained from different calibrations for PantheonPlus and DES Dovekie, respectively. This makes the $\eta_\text{obs}$ values derived from DES Dovekie (blue markers) significantly lower than those from PantheonPlus (red markers) at the third BAO redshift $z=0.934$ in Figure \ref{fig:eta errobar}.
In any case, more $\eta_\text{obs}$ data are required for a decisive answer to this discrepancy.

In sum, the results obtained in this section using Method II are generally consistent with those from the previous section, although the smaller number of data points leads to some differences. Despite this, Method II has the advantage of allowing $\eta_\text{obs}$ to be directly constructed from the data without assuming any parametrization of $\eta(z)$.
The obvious shortcoming of the present analysis can be overcome by incorporating more data on $d_A$ within the redshift range covered by the current SN data, or by extending the SN redshift range to include more BAO data.

	\begin{table*}[tbp]
	\setlength\tabcolsep{3pt}
	\renewcommand{\arraystretch}{1.5}
	\centering
	\caption{$\eta_0$ for different calibrations and SN data sets. Similar to Table \ref{tab: calibrations PP} and \ref{tab: calibrations DES}, $D_\text{com}$=SN+BAO, where the SN data sets are PantheonPlus (left column) and DES Dovekie (right column). }
	\label{tab:eta0 physical zeta}
	\vspace{3mm}
	\scalebox{0.85}{
		\begin{tabular}{llccccccccc}
			\hline
			\multirow{2}{*}{$\eta(z)$} & \multirow{2}{*}{ data set}  & \multicolumn{2}{c}{PantheonPlus} &  \multicolumn{2}{c}{DES Dovekie} \\ 
			& & $\eta_0$ & $\zeta$ & $\eta_0$ & $\zeta$\\
			\midrule[0.4pt]
			\multirow{3}{*}{P1} 
			& D$_\text{com}$+$M_B$ & $-0.010 \pm 0.013$ & $51.9\pm 1.1$ & $-0.021\pm 0.019$ & $51.1\pm1.1$\\
			& D$_\text{com}$+CC & $-0.010 \pm 0.022$ & $51.9\pm2.3$ & $-0.021 \pm 0.037$ & $51.2\pm2.3$
			\\ 
			& D$_\text{com}$+$r_d$ & $-0.011 \pm 0.010$ & $51.85\pm 0.65$ &$-0.021 \pm 0.014$  & $51.07\pm0.69$\\  
			
			\midrule[0.4pt]
			\multirow{3}{*}{P2} 
			& D$_\text{com}$+$M_B$ & $-0.032 \pm 0.028$ & $51.7\pm1.2$ & $-0.051 \pm 0.035$ & $50.9\pm 1.2$\\
			& D$_\text{com}$+CC & $-0.028 \pm 0.052$ & $51.7\pm2.6$  & $-0.048 \pm 0.072$ & $50.9\pm 2.5$
			\\ 
			& D$_\text{com}$+$r_d$ & $-0.031 \pm 0.023$ & $51.69\pm 0.84$ & $-0.052 \pm 0.028$ & $50.84\pm 0.87$\\  
			
			\midrule[0.4pt]
			\multirow{3}{*}{P3} 
			& D$_\text{com}$+$M_B$ & $-0.019 \pm 0.019$ & $51.8\pm1.2$ &$-0.032 \pm 0.026$ & $51.0\pm 1.2$\\
			& D$_\text{com}$+CC & $-0.018 \pm 0.033$ & $51.8\pm 2.4$  &$-0.031 \pm 0.053$ & $51.0\pm 2.5$
			\\ 
			& D$_\text{com}$+$r_d$ & $-0.020 \pm 0.016$ & $51.73\pm0.79$ & $-0.032 \pm 0.021$ & $51.00\pm0.81$\\  
			
			\hline
		\end{tabular}
	}
\end{table*}

\begin{figure*}[tbp]
	\centering
	\includegraphics[scale=0.7]{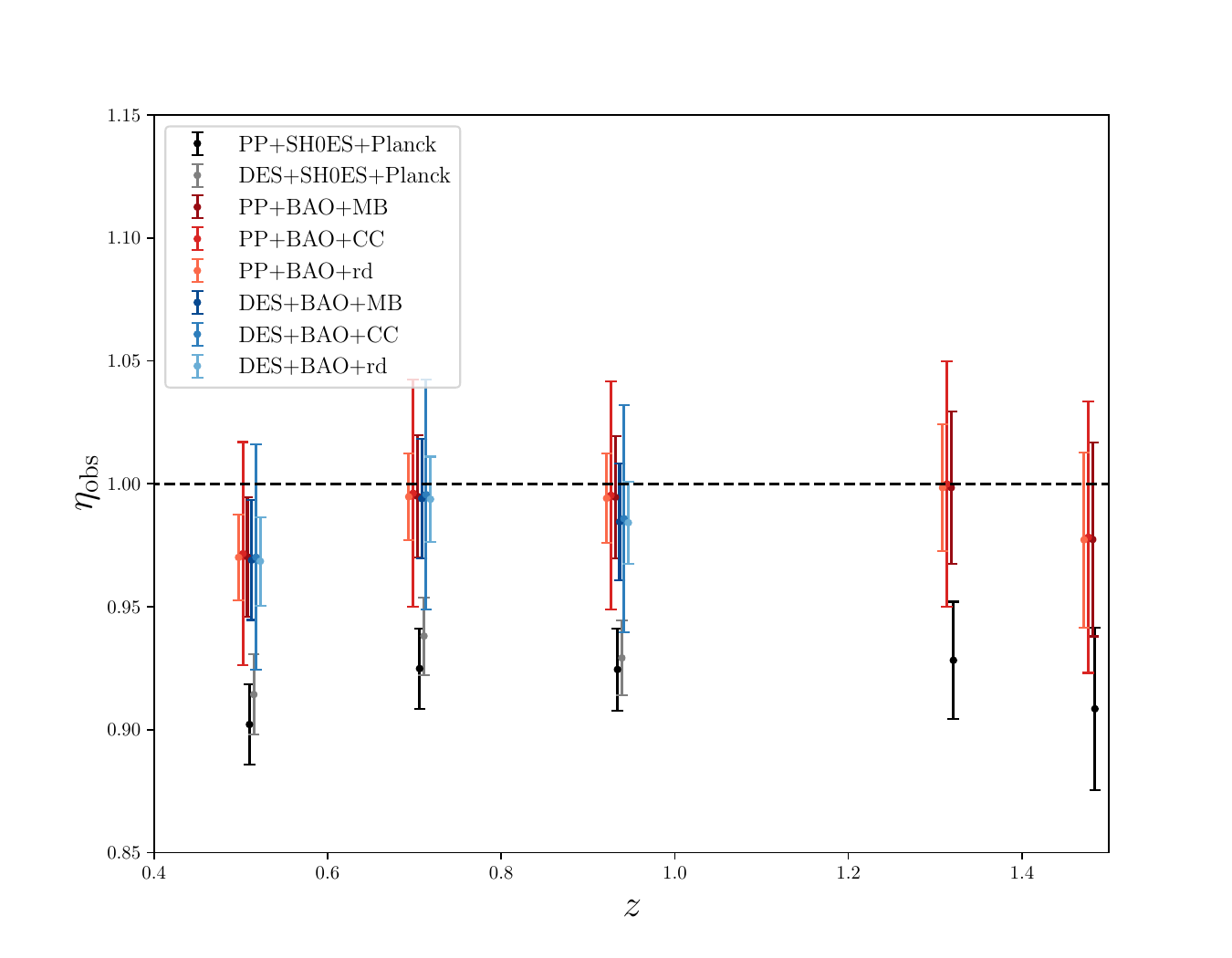} 
	\caption{Constructed $\eta_\text{obs}$ with $1\sigma$ error bars using various calibrations at five $z_\text{BAO}$. To make the numerous error bars at the same $z_\text{BAO}$ more distinct, the markers are slightly shifted by $0.005$ in redshift relative to one another.  The black and grey markers correspond to $\eta_\text{obs}$ constructed with $\zeta$ derived from the SH0ES $M_B$ and Planck $r_d$. Various kinds of red colors and blue colors denote the results obtained using PantheonPlus and DES Dovekie, respectively, with $r_d$ and $M_B$ values from the three calibrations listed in Tables \ref{tab: calibrations PP} and \ref{tab: calibrations DES}. The black dashed horizontal line corresponds to $\eta_\text{obs}=1$.   }
	\label{fig:eta errobar}
\end{figure*}

\section{Conclusions}

This paper presents a model-independent test of the CDDR using two complementary methods. Method I employs the PAge parametrization to jointly constrain cosmological parameters and parametric forms of CDDR violation, using various data including SN (PantheonPlus and DES Dovekie), BAO (DESI DR2), CC, and GRB. The GRB data, while extending the redshift range to $z \sim 8$, are found to have limited constraining power compared to the other data, due to the large intrinsic scatter in the Amati relation. Therefore they do not significantly improve the results. The analysis shows that the CDDR violation parameter $\eta_0$ is consistent with zero at the $1\sigma$ level for all parametrizations considered. Moreover, the different calibrations embodying the distance ladder (SN+BAO+$M_B$ SH0ES prior) and the inverse distance ladder (SN+BAO+CC and SN+BAO+$r_d$ Planck prior) yield consistent results of no CDDR violation. Notably, when the SH0ES \(M_B\) and Planck \(r_d\) priors are imposed simultaneously, a spurious violation of the CDDR appears at the $3\sigma\sim 4\sigma$ level, reflecting the well-known tension between early-time and late-time measurements rather than genuine physics violating the CDDR. Throughout the analysis, the two SN data sets, PantheonPlus and DES Dovekie, produce consistent results in the test of the CDDR.

Method II uses the GP method to reconstruct the luminosity distance from SN data and then constructs $\eta_\text{obs}$ at BAO redshifts. No \textit{a priori} parametrization $\eta(z)$ of the CDDR violation is required in this method, thereby introducing no possible bias compared to Method I. This approach yields consistent results with Method I, again showing no evidence for CDDR violation. Overall, both methods confirm the validity of the CDDR within current observational uncertainties, while highlighting the importance of calibration choices. 

Further model-independent and non-parametric tests of the CDDR will benefit significantly from future high-quality data with increased statistics and wider redshift coverage. In particular, more precise and abundant data of $d_L$ and $d_A$ are expected from forthcoming observations from facilities such as the Roman Space Telescope \cite{1503.03757}, the Rubin Observatory \cite{0805.2366}, Euclid \cite{1110.3193}, and the China Space Station Telescope \cite{2507.04618}, which will  lead to more stringent and reliable tests of the CDDR. 
Moreover, the extension of CDDR tests to high redshifts using GRB data will further rely on improved understanding of GRB physics and better control of their systematics, as well as larger data samples from future missions such as THESEUS \cite{2104.09531} and SVOM \cite{2203.10962}. In addition, incorporating emerging probes \cite{2201.07241} such as strong gravitational lensing, radio quasars, and gravitational waves will further enhance the redshift coverage and increase the number of independent distance measurements \cite{1511.01318,2110.00927,2512.06454,2602.16869}. These directions will be explored in future work.

\appendix
\section{Constant CDDR violation as a recalibration of $M_B$}
\label{app:const eta}
 
For a constant $\eta(z)=1+\eta_0$, adding $\eta(z)$ is equivalent to a recalibration of $M_B\rightarrow\tilde M_B$, and the parameter $\eta_0$ is degenerate with $M_B$, as can be easily seen from $m_B-M_B=5\log_{10}[\eta(1+z)^2d_M] +25$, and the degenerate direction in the $\eta_0$-$M_B$ plane is given by 
\be \label{degeneracy MB eta0}
M_B+5\log_{10}(1+\eta_0)\equiv \tilde M_B=const.
\ee 
For small $\eta_0$, this can be approximated by a linear relation 
\be \label{degeneracy MB eta0 approx}
M_B+\frac{5}{\ln 10}\eta_0 =\tilde M_B.
\ee 
The numerical result can clearly exhibits this degeneracy. To illustrate this point, consider the result of applying PantheonPlus+BAO+$r_d$ prior, shown in Figure \ref{Fig:CDDR const}. As expected, $M_B$ and $\eta_0$ are not well constrained (although the combination $\tilde M_B$ is in fact well constrained). The exact degenerate direction is given by the solid curve (orange), corresponding to $\tilde M_B=-19.40$, while the linear approximation for small $\eta_0$ is given by the dashed straight line (red).     
\begin{figure*}[tbp]
	\centering
	\includegraphics[scale=0.8]{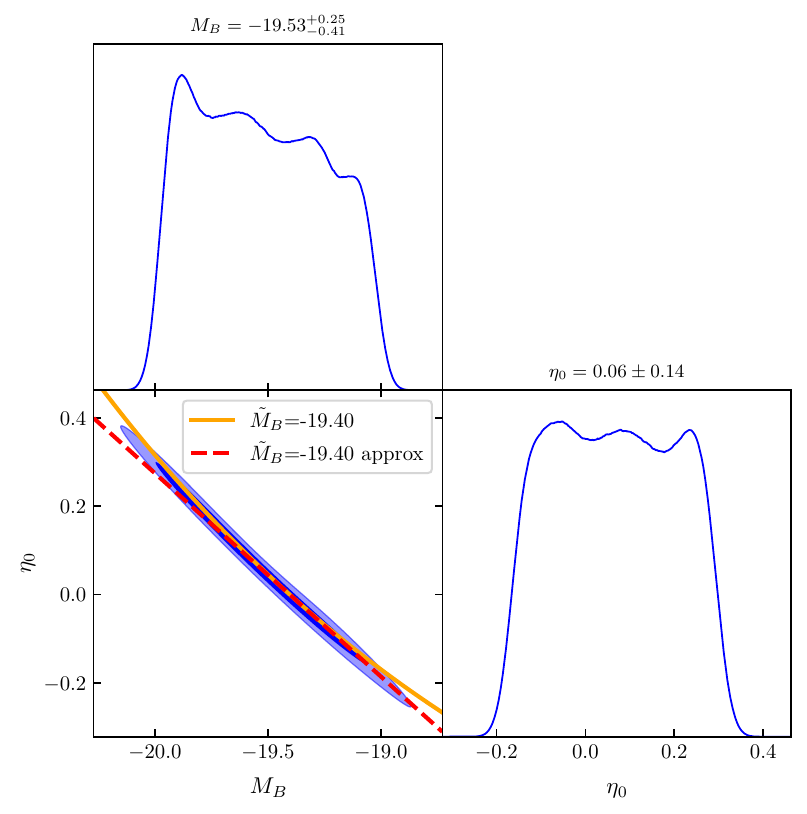} 
	\caption{Illustration of the degeneracy of $M_B$ and $\eta_0$. The orange solid curve represents \eq{degeneracy MB eta0} with $\tilde M_B=-19.40$, and the red dashed straight line represents the linear approximation \eq{degeneracy MB eta0 approx}.  }
	\label{Fig:CDDR const}
\end{figure*}

\section{Effect of large intrinsic scatter in the Amati relation}
\label{app:A123}

To further illustrate the effect of large intrinsic scatter in the Amati relation, we consider using a different GRB data set from the 15-year Fermi-GBM catalog recently compiled in \cite{2405.14357}, denoted as A123. The results for the linear $\eta(z)$ plotted in Figure \ref{fig: A123 comparison} show that the constraints are also rather weak compared to that of SN+BAO+CC, due to its large intrinsic scatter $\sigma_\text{int}\sim 0.5$. Moreover, the results involving $p_{age}$ of A123 alone exhibit deviations from those of SN+BAO+CC much larger than $2\sigma$. This is similar to the observation in \cite{2502.08429} that the constraints from A123 on the matter parameter $\Omega_m$ in $\Lambda$CDM and $\phi$CDM there exhibit a tension larger than $2\sigma$  with those from BAO+CC data. Of course, A123 by itself is internally consistent and still worthy of further investigation; it is just that this data set should not be combined with BAO+CC.

\begin{figure*}[tbp]
	\centering
	\includegraphics[scale=0.8]{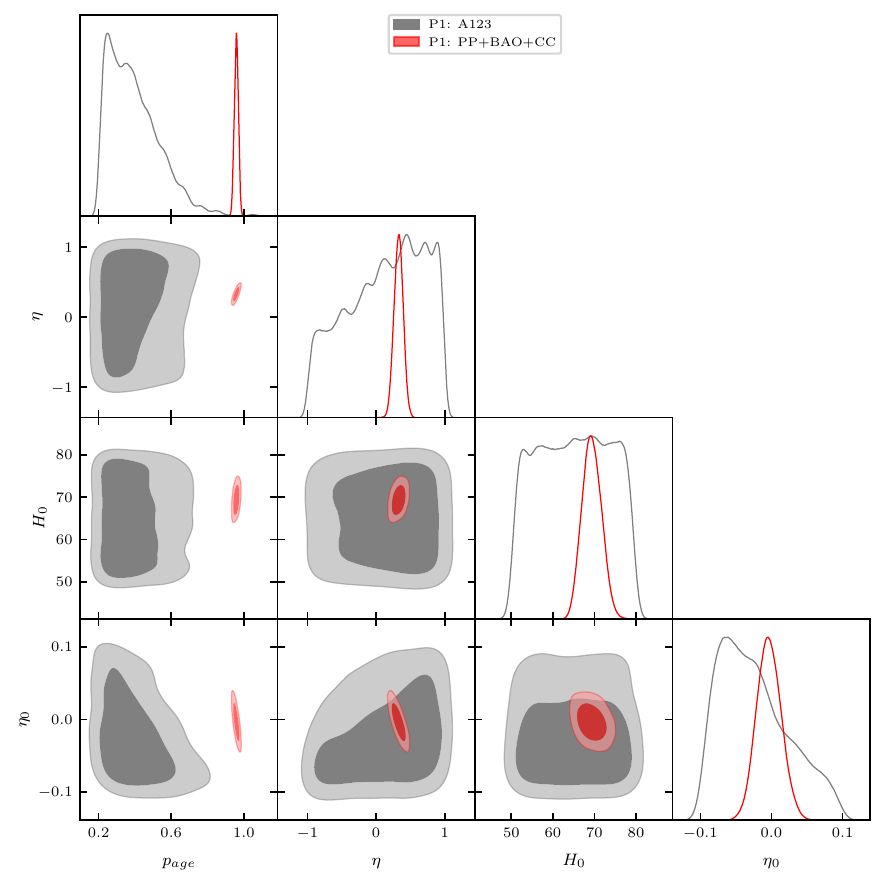} 
	\caption{Illustration of the weak constraints of GRB data set A123 (grey) compared to PantheonPlus+BAO+CC (red), in the linear $\eta(z)$ case of P1.  }
	\label{fig: A123 comparison}
\end{figure*}




\section{$a_B$ tension between PantheonPlus and DES Dovekie}
\label{app: aB}
 
When using the SN data, the degenerate parameters $M_B$ and $H_0$ can be combined into one parameter $a_B$, such that $-5a_B$ corresponds to the intercept in the relation
\be 
m_B=5\log_{10}D_L-5a_B,
\ee
where
\be 
-5a_B\equiv M_B+5\log_{10}\frac{c}{\text{Mpc} H_0}+25,
\ee
and $D_L\equiv d_LH_0/c$ is the dimensionless luminosity distance independent of $H_0$. 

We consider the linear $\eta(z)$ of case P1 as an example. The conclusion holds for the other parametrizations of $\eta(z)$. The results from PantheonPlus, DES Dovekie, and the previous DES5yr (for comparison), are listed in Table \ref{tab: aB tension}, and plotted in Figure \ref{fig: aB tension}. The $H_0$ values are all essentially the same. The $M_B$ and $\eta_0$ values show noticeable differences, though they are mutually compatible within $1\sigma$. However, the $a_B$ values show significant discrepancy. In particular, $a_B$ from DES Dovekie and DES5yr deviate from that of PantheonPlus by $\sim3\sigma$ and $\sim5\sigma$, respectively.  These observations show that DES Dovekie produces results closer to those of PantheonPlus than the previous DES5yr does. 
	
	\begin{table*}[tbp]
		\setlength\tabcolsep{3pt}
		\renewcommand{\arraystretch}{1.5}
		\centering
		\caption{ Illustration of constraints using different SN data sets, in the linear $\eta(z)$ case of P1. }
		\vspace{3mm}
		\label{tab: aB tension}
		\scalebox{0.85}{
			\begin{tabular}{llccccccccc}
				\hline
				data set &  $M_B$ & $H_0$  & $a_B$ & $\eta_0$ \\
				\midrule[0.4pt]
				PantheonPlus & $-19.373\pm0.071$ & $69.4\pm2.3$ &$ -4.7614\pm 0.0014$  & $-0.004\pm0.017$ \\
				
				\midrule[0.4pt]
				DES Dovekie & $-19.342\pm 0.072$ & $69.3^{+2.2}_{-2.5}$ & $-4.7679\pm0.0021$ & $-0.003\pm0.017$ \\
				
				\midrule[0.4pt]
				DES5yr & $-19.325\pm 0.071$ & $69.3\pm2.3$ & $-4.7712\pm 0.0022$  & $-0.014\pm0.018$ \\
				\hline
			\end{tabular}
		}
	\end{table*}
	
	\begin{figure*}[tbp]
		\centering
		\includegraphics[scale=0.5]{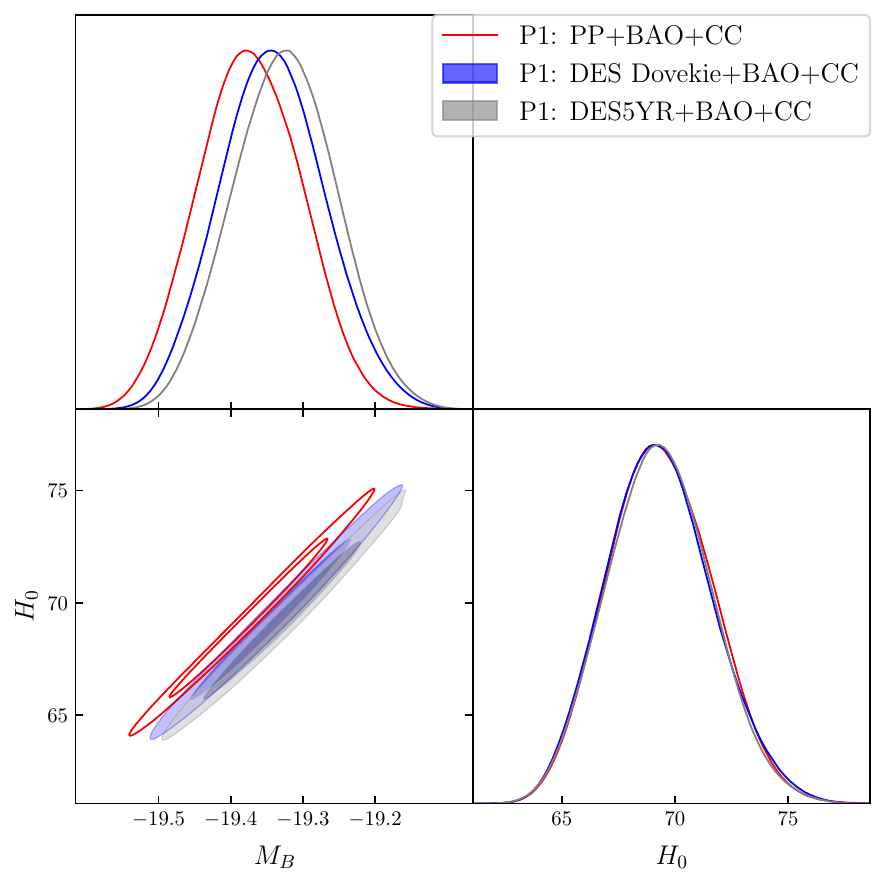} 
		\includegraphics[scale=0.5]{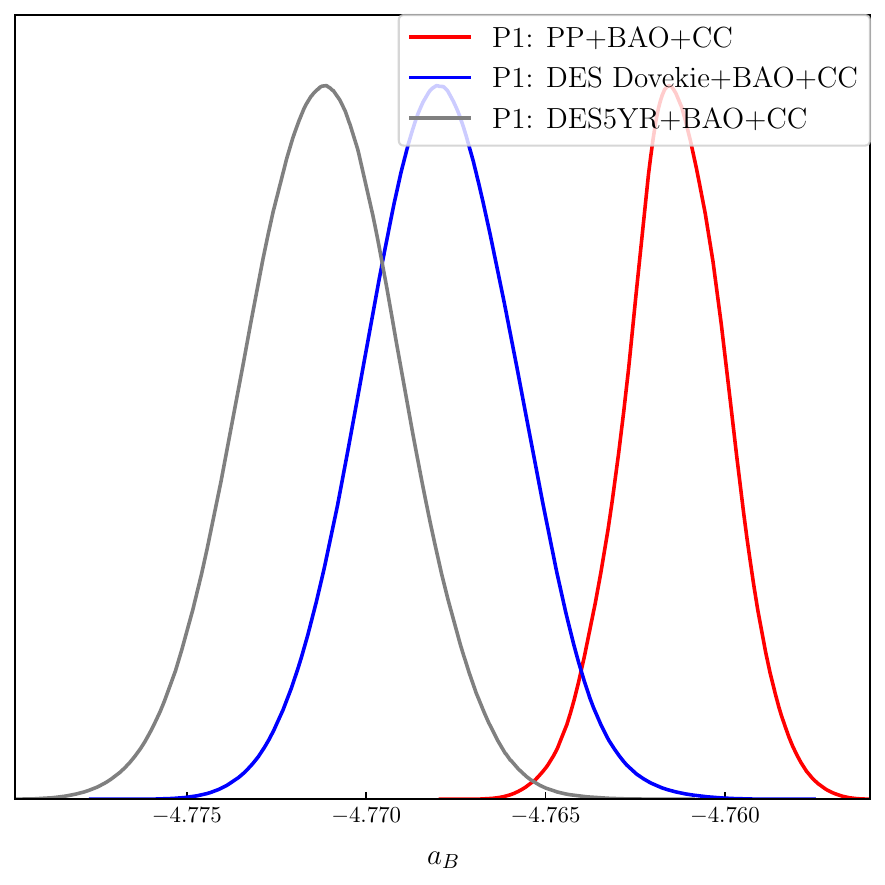} 
		\caption{Comparison of the results from PantheonPlus (red), DES Dovekie (blue), and DES5yr (grey), in the linear $\eta(z)$ case of P1. }
		\label{fig: aB tension}
	\end{figure*}

\section{GP reconstruction for $m_B$}
\label{app:GP for mB}
 
The GP reconstruction of $m_B$ (or $\mu$) may exhibit unphysical wiggles, as has been noticed before (e.g., \cite{2206.15081}, \cite{2211.02473}, and \cite{2509.19899}). But, if one first transforms the $m_B$ data into an exponential form, such as $10^{0.2 m_B-5}$, as is common in the literature of GP reconstruction, the wiggles disappear. The wiggles can be found in $m_B(z)$ (or $\mu(z)$) reconstructions from  Pantheon \cite{1710.00845}, PantheonPlus and Union3 \cite{2311.12098}, where the data points are dense at $z\lesssim 1$ and sparse at higher redshifts. For DES Dovekie, no significant wiggles appear in the $m_B$ reconstruction, as shown in the left panel of Figure~\ref{fig:GP Dovekie}. 

\begin{figure*}[tbp]
	\centering
	\includegraphics[scale=0.33]{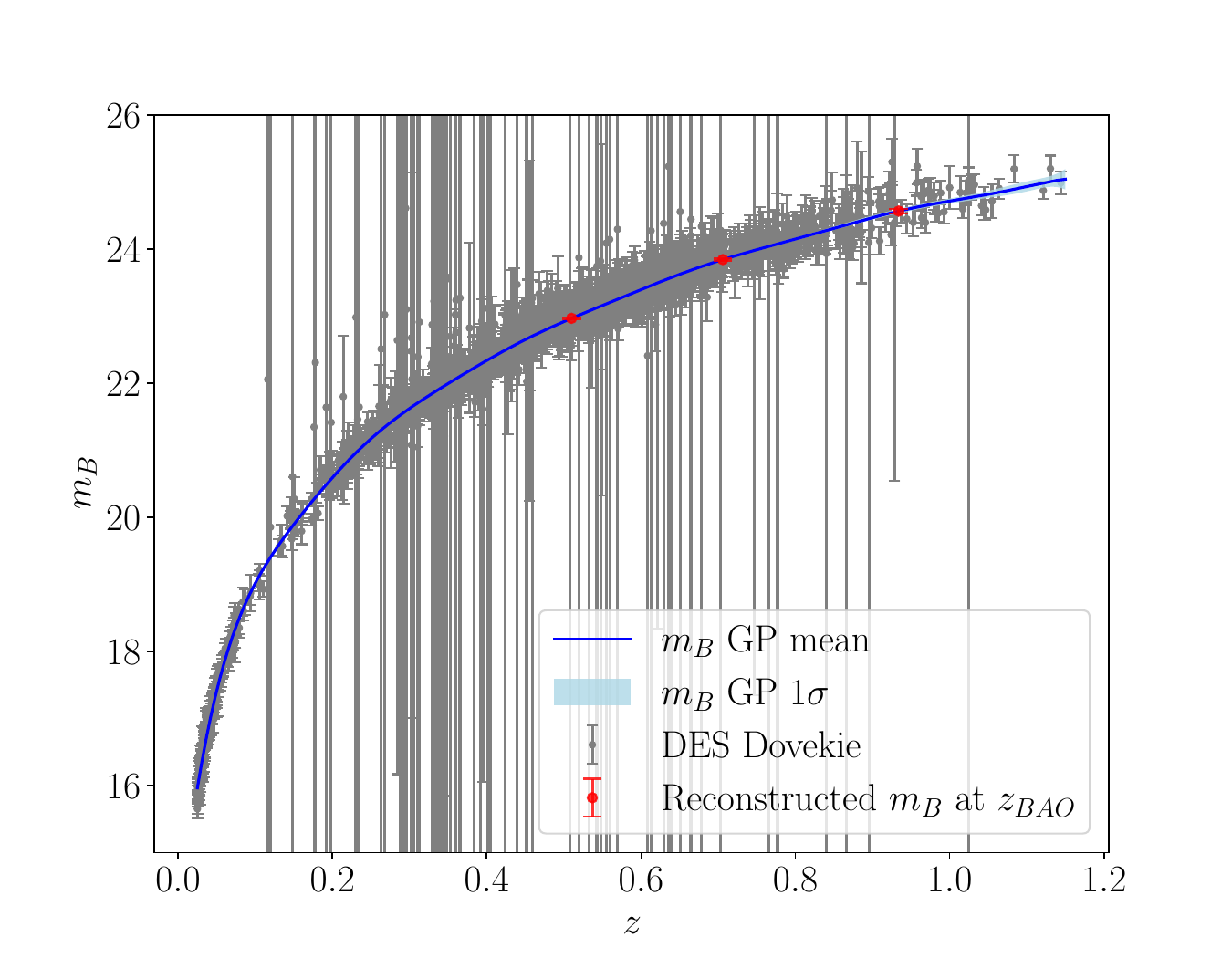} 
	\includegraphics[scale=0.33]{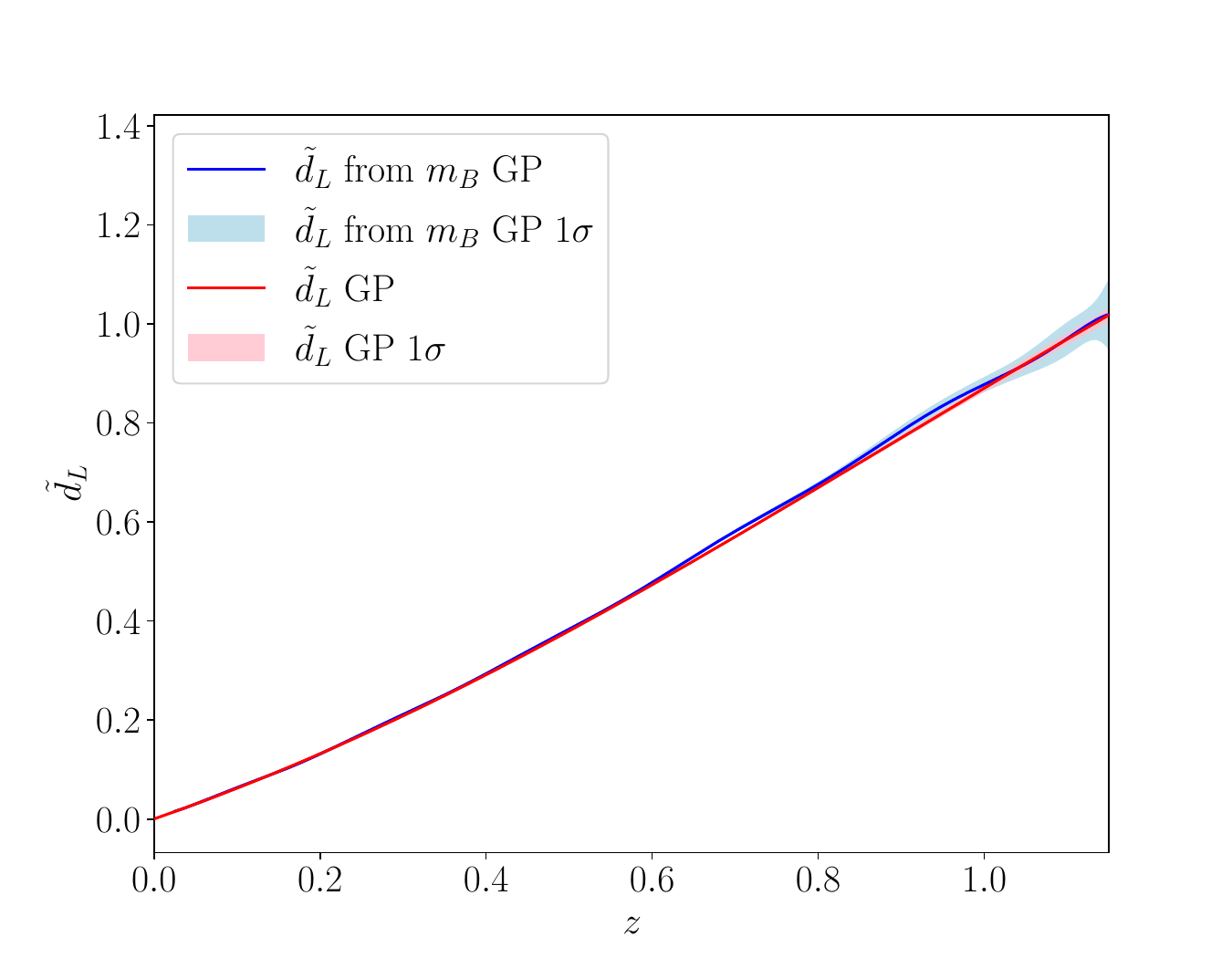} 
	\caption{Different GP reconstructions using DES Dovekie. Left panel: GP reconstruction of $m_B(z)$ from the $m^{corr}_B$ data. Right panel: comparison of $\tilde{d}_L$ constructed from the GP reconstruction of $m_B$ (blue), and the GP reconstruction of $\tilde d_L$ directly from the transformed data (red).}
	\label{fig:GP Dovekie}
\end{figure*}

The correlation length $l$ of the hyperparameters determines the scale over which significant variations occur. For Pantheon and PantheonPlus, $l\sim  0.14$. For Union3, $l\sim 0.3$. For DES Dovekie, $l\sim 0.19$. 

Among all these data sets, the data points of DES Dovekie are distributed more uniformly within the redshift range $z\in(0,1.1)$, while the other three are sparse at $z\gtrsim1$. It is in this higher redshift sparse range that the wiggles appear. This suggests that the range with dense data points dominates the GP training, resulting in a small correlation length $l\sim 0.1$ to $0.3$. For the sparse range, however, the separation between neighboring points is larger than this $l$, thus causing the wiggles. 
 
The next question is why the transformed data, e.g., $\tilde d_{L}$ or $d_M$ ( which are proportional to $10^{0.2 m_B}$), yield larger correlation lengths ($l\sim 1.4$ or $2.2$ for PantheonPlus). The reason can be understood as follows. $m_B$ (or $\mu$) is essentially the logarithm of distance. The distance is roughly a power-law function (or polynomial) of $z$. Thus, $m_B$ (or $\mu$) is roughly a logarithmic function of $z$. For data points which concentrate on the range $z\lesssim 1$, the variations of the neighboring data points are significant. If the number of data points in this range dominates, as is the case for most current SN data sets, the training of the hyperparameters yileds a small correlation length $l$. When this $l$ is used to make predictions in the sparse range $z\gtrsim 1$, wiggles arise, as mentioned above. Indeed, the wiggles can be removed by adding denser data points in this range. This is the reason why DES Dovekie has the same correlation length $l\sim 0.1$ to $0.2$ as the other data sets, yet shows almost no wiggles. Again, DES Dovekie has almost uniform distribution of data points throughout $z\in(0,1.2)$. 
However, after the exponential transformation $\sim 10^{0.2m_B}$, the transformed data approximately follow a power law in $z$, making the variations more uniform across the full range. This leads to larger correlation length $l\sim 1$ to $2$, reducing the wiggles and producing smooth plots.

\acknowledgments
The author thanks Yun Chen for correspondence regarding the PantheonPlus data, and Brodie Popovic for correspondence regarding the DES Dovekie data.
	
	\bibliographystyle{JHEP}   
	\bibliography{references}  
	
\end{document}